\documentclass[fleqn,usenatbib]{mnras}
\pdfoutput=1 
\usepackage{newtxtext,newtxmath}

\usepackage[T1]{fontenc}

\usepackage{graphicx}	

\newcommand{\hmpc}{h^{-1}{\rm Mpc}}
\newcommand{\hgpc}{h^{-1}{\rm Gpc}}

\newcommand{\mbr}{\mathbf r}
\newcommand{\cdf}{{\rm CDF}}
\newcommand{\pdf}{{\rm PDF}}
\newcommand{\nn}{{\rm NN}}

\newcommand{\second}{$2$nd}

\newcommand{\eq}[2]{\begin{align} \label{eq:#1} #2 \end{align}}

\title[Cosmology with Nearest Neighbor Distributions]{Nearest Neighbor distributions: new statistical measures for cosmological clustering}

\author[Banerjee \& Abel]{
Arka Banerjee\thanks{E-mail: {\tt arkab@stanford.edu}}
and Tom Abel\thanks{E-mail: {\tt tabel@stanford.edu}} \\
Kavli Institute for Particle Astrophysics and Cosmology, Stanford University, 452 Lomita Mall, Stanford, CA 94305, USA \\
Department of Physics, Stanford University, 382 Via Pueblo Mall, Stanford, CA 94305, USA \\
SLAC National Accelerator Laboratory, 2575 Sand Hill Road, Menlo Park, CA  94025, USA
}

\date{Accepted XXX. Received YYY; in original form ZZZ}

\pubyear{2020}
\begin{document}
\label{firstpage}
\pagerange{\pageref{firstpage}--\pageref{lastpage}}
\maketitle

\begin{abstract}
The use of summary statistics beyond the two-point correlation function to analyze the non-Gaussian clustering on small scales, and thereby, increasing the sensitivity to the underlying cosmological parameters, is an active field of research in cosmology. In this paper, we explore a set of new summary statistics --- the $k$-Nearest Neighbor Cumulative Distribution Functions ($k\nn$-$\cdf$). This is the empirical cumulative distribution function of distances from a set of volume-filling, Poisson distributed random points to the $k$--nearest data points, and is sensitive to all connected $N$-point correlations in the data. The $k\nn$-$\cdf$ can be used to measure counts in cell, void probability distributions and higher $N$--point correlation functions, all using the same formalism exploiting fast searches with spatial tree data structures. We demonstrate how it can be computed efficiently from various data sets - both discrete points, and the generalization for continuous fields. We use data from a large suite of $N$-body simulations to explore the sensitivity of this new statistic to various cosmological parameters, compared to the two-point correlation function, while using the same range of scales. We demonstrate that the use of $k\nn$-$\cdf$ improves the constraints on the cosmological parameters by more than a factor of $2$ when applied to the clustering of dark matter in the range of scales between $10\hmpc$ and $40\hmpc$. We also show that relative improvement is even greater when applied on the same scales to the clustering of halos in the simulations at a fixed number density, both in real space, as well as in redshift space. Since the $k\nn$-$\cdf$ are sensitive to all higher order connected correlation functions in the data, the gains over traditional two-point analyses are expected to grow as progressively smaller scales are included in the analysis of cosmological data, provided the higher order correlation functions are sensitive to cosmology on the scales of interest.
\end{abstract}




\section{Introduction}
\label{sec:intro}

Over the past three decades, a large part of the progress in cosmology --- especially in the pursuit of increasingly precise and accurate constraints on the standard cosmological paramemeters --- has been predicated on the use of two-point statistics in the analysis of cosmological datasets, either in real space, or in Fourier space. This includes analysis of the Cosmic Microwave Background \citep{2018arXiv180706209P,2013ApJS..208...19H} as well as Large Scale Structure analyses at low redshifts \citep{2017MNRAS.470.2617A,2020Ivanov,2020JCAP...05..005D,2017MNRAS.465.1454H,2018PhRvD..98d3526A}. The latter include the analysis of both galaxy clustering --- discrete data points --- and weak lensing measurements --- in the form of continuous maps. These studies employ a wide range of theoretical approaches in their modeling, including linear perturbation theory, higher order perturbation theory, and $N$-body simulations, but the two-point correlation function, or the power spectrum is usually the summary statistic of choice to compare theoretical predictions and data. Given its widespread use, a number of tools have been developed over the years for fast and efficient calculation of two-point statistics from cosmological data, or from simulations.

The two-point statistics provides a complete statistical description of a Gaussian random field. Perturbations in the early Universe are believed to closely follow the statistics of a Gaussian random field \citep[\textit{e.g.}][]{peacock_1998}, though attempts have also been made to quantify any departures from Gaussianity in the early Universe \citep{2019arXiv190505697P}. As long as the evolution of the early Universe perturbations remain well described by linear perturbation theory, the Gaussian nature of the field is not affected. High redshift CMB analyses $(z\sim 1100)$ and the analyses of clustering on extremely large scales at low redshift fall in this regime. Therefore, current analyses, utilizing two-point statistics, are able to extract maximal information about cosmological parameters from these redshifts and scales. On the other hand, at low redshifts, the field develops non-Gaussian features sourced by continued gravitational collapse --- smaller the scale, more non-Gaussian the underlying field. Higher connected $N$-point correlation functions, which encode the more complicated nature of the field, start to be statistically important. The two-point statistics themselves can still be accurately modeled on relatively small scales, either through higher order perturbation theories \citep{2012JHEP...09..082C,2012PhRvD..86j3528T,2015PhRvD..91b3508V, 2020JCAP...05..005D,2020Ivanov}, or using $N$-body simulations \citep{2015MNRAS.454.1958M,2019MNRAS.484.5509E,2017ApJ...847...50L,2017PhRvD..96l3515N}. Nonetheless, an analysis using only the two-point statistic  is insufficient to probe all the effects of different cosmological parameters on the evolution of the overall cosmological field. This motivates the need for considering other summary statistics in the analysis of clustering on smaller scales. Since the total information available in a survey scales with the number of independent modes \citep{1997PhRvL..79.3806T}, and given that there are many more independent modes on small scales, it is important to develop these new statistical methods, that better extract information from small scales, to make optimal use of data from ongoing and future cosmological surveys.

Various approaches toward harnessing information beyond that contained in the two-point statistics have been explored in the literature. One method is consider the higher $N$-point functions in the analysis --- the three-point function and its Fourier transform, the bispectrum \cite[\textit{e.g.}][]{1998ApJ...496..586S,2004MNRAS.348..897T,2006PhRvD..74b3522S}, or the four-point function and its Fourier transform, the trispectrum \cite[\textit{e.g.}][]{2001ApJ...553...14V}. While these additional statistics can be highly informative about small scale clustering, and promise to yield significantly tighter constraints on some cosmological parameters \citep{2020JCAP...03..040H,2019JCAP...05..043C}, one drawback is that they are generally entail much higher computational costs to measure either from simulations, or from data. The computational complexity rises with $N$ - the order of the highest connected correlation function considered, while keeping the data size fixed. Further, the noise properties for these higher order estimators often make it difficult to obtain good signal-to-noise ratio (SNR) over a wide range of scales. In spite of these issues, analyses including higher order correlations, especially the bispectrum, have successfully been applied to certain cosmological datasets \citep{2015MNRAS.451..539G,2019MNRAS.484.3713G,2017MNRAS.469.1738S}. 

In this context, there also exist approaches which attempt to undo the nonlinear effects of gravitational clustering on various cosmological fields. Once this procedure is applied, the analysis proceeds with the measurement of the two-point function of the linearized field. This approach has already been applied specifically to the reconstruction of the Baryon Acoustic Oscillation (BAO) signal \citep{Eisenstein_2007,PhysRevD.79.063523,2012MNRAS.427.2132P}. \citet{2015PhRvD..92l3522S} showed that the reconstruction process can be interpreted as the transfer of information from the higher connected $N$-point functions in the observed field into the reconstructed two-point function. More recently, methods have been proposed for the full reconstruction of the initial linear modes from nonlinear observables at low redshift \citep{PhysRevD.81.063531,2017JCAP...12..009S,2019JCAP...10..035H,2018JCAP...10..028M}. 

There is also a large body of literature investigating the one point Probability Distribution Function (PDF) of matter density in the Universe, and the related counts-in-cell (CIC) statistics for discrete tracers like dark matter halos or galaxies \citep{1991MNRAS.248....1C,1994ApJ...435..536C,1994ApJ...420...44K,2000ApJ...539..522G,2008MNRAS.386..407L,PhysRevD.90.103519,10.1093/mnras/stw1074,2018MNRAS.481.4588K}. While the PDF or CIC statistics are close to Gaussian when evaluated on large scales, and therefore contain the same information as the two-point function, on small scales the PDF captures information about all higher moments of the distribution, and therefore can be used to place stronger constraints on various cosmological parameters \citep{2019arXiv191111158U}. Particular attention has been paid to the case of a lognormal density PDF and the Gaussianization of the density field through logarithmic transforms \cite{2011ApJ...738...86C,2011ApJ...742...91N,2012ApJ...750...28C,2014MNRAS.439L..11C}. Variations of this type of analysis have also been applied already to different cosmological datasets \citep{PhysRevD.91.103511,10.1093/mnras/stv2506,PhysRevD.98.023507,PhysRevD.98.023508,2020arXiv200601146R}. While this approach is extremely attractive in terms of its sensitivity to all higher order correlations, calculating the PDF from data involves multiple steps of smoothing and averaging, and the calculations have to be done separately for every radius bin used in the analysis.

Other statistical measures that have been employed to extract non-Gaussian information from cosmological fields, especially in the context of weak lensing, include peak counts \citep{2017A&A...599A..79P,2018JCAP...10..051F} and Minkowski functionals \citep{2010PhRvD..81h3505M,2012MNRAS.419..536M,2013PhRvD..88l3002P}. Yet another set of studies have attempted to use special properties of nonlinear regions, such as halos and voids, in specific cosmologies, to enhance the constraints on some of the cosmological parameters. These include searches for scale dependent bias on large scales in the context of massive neutrinos \citep{2014JCAP...03..011V,2016PhRvD..93j3526L,2016JCAP...11..015B,2019PhRvL.122d1302C,2019arXiv190706598B} and primordial non-Gaussianity \citep{PhysRevD.77.123514,2009MNRAS.396...85D,2018PhRvL.121j1301C}, and the use of marks, or density dependent weights, for correlation functions in the context of modified gravity \citep{2016JCAP...11..057W, 2018MNRAS.479.4824H, 10.1093/mnras/sty1335} and massive neutrinos \citep{2020arXiv200111024M}. While these methods attempt to use additional information from the nonlinear density field, the statistic considered is usually the two-point function.

Finally, there are studies exploring the clustering of halos or galaxies in terms of the Void Probability Function (VPF), which was also shown to be the generating function for the full distribution of the clustering, and sensitive to all connected $N$-point functions \cite{White1979,1986ApJ...306..358F,10.1093/mnras/stt745}. A related approach is explored in \citet{2020MNRAS.495.3233P}. VPF and related measurements have already been applied to data \citep{Sharp:1981aa,2011ApJ...727...48W,2019MNRAS.488..470W}. Over the years, the concept of the generating function, beyond just probabilities of finding completely empty volumes, as captured in the VPF, was further developed in the context of cosmological clustering \citep{Balian1989,Szapudi1993,1994A&A...291..697B}, and provides an overarching theoretical framework to connect the parallel approaches of using higher $N$-point functions, and the use of one point PDF analysis, toward the extraction of non-Gaussian information on small scales.

In this paper, we introduce the $k$-nearest neighbor Cumulative Distribution Functions ($k\nn$-$\cdf$), \textit{i.e.}, the empirical cumulative distribution function of distances from a set of volume-filling Poisson distributed random points to the $k$--nearest data points. The $k\nn$-$\cdf$ are a set of new summary statistics that can be applied to the clustering analysis of cosmological datasets - both discrete tracers, and continuous fields, where the latter can be sampled by a set of tracers. We set out the connections between these new statistics and the generating function formalism. Through the latter, we describe the relationship between the $k\nn$-$\cdf$ statistics and the statistics of higher $N$-point functions, as well as the density PDF over a range of scales. Importantly, we demonstrate how $k\nn$-$\cdf$ statistics can be computed efficiently on a given dataset - and how a single measurement step is sufficient to gain information about all $N$-point correlation functions present in the data over a relatively broad range of scales. We apply these statistics in the context of familiar distributions, and finally, quantify the improvements in cosmological parameter constraints, compared to two-point function analyses over the same range of scales.

The layout of the paper is as follows: in Sec. \ref{sec:NNCDF}, we introduce the mathematical framework relevant for the $k\nn$-$\cdf$ statistics, and outline how they can be computed for a given dataset. In Sec. \ref{sec:NNCDF_examples}, we apply the $k\nn$-$\cdf$ statistics to various underlying fields to illustrate its novel features. In Sec. \ref{sec:cosmo_constraints}, the constraints on cosmological parameters using the $k\nn$-$\cdf$ statistics is explored. Finally, in Sec. \ref{sec:conclusions}, we summarize our main findings, and discuss possible directions in which this study can be extended.

\section{Introduction to Nearest Neighbor Cumulative Distribution Functions}
\label{sec:NNCDF}
In this section, we introduce the concept of the Nearest Neighbor Cumulative Distribution Function, and explore its connections to other statistical measures for clustering used in the literature. We also outline the method by which $\nn$-$\cdf$ can be computed quickly for a given dataset.

\subsection{Formalism}
\label{sec:formalism}

We consider a set of tracers of a underlying continuous field, with mean number density $\bar n$ and connected $N$-point correlation functions denoted by $\xi^{(N)}$. $\xi^{(0)}=0$ by definition, and $\xi^{(1)}=1$ to correctly normalize the distribution. The generating function, $P(z|V)$, of the distribution of the counts of data points enclosed in volume $V$ can be written as \citep{White1979, Balian1989,Szapudi1993}:
\begin{align}
	P(z|V) &= \sum_{k=0}^{\infty} P_{k|V} z^k \nonumber\\
	&= \exp\Bigg[\sum_{k=1}^\infty \frac{\bar n ^k (z-1)^k }{k!} \times  \nonumber \\ 
	& \quad \qquad \int _V ... \int_V d^3 \mathbf r_1 ... d^3 \mathbf r_k \xi^{(k)} (\mathbf r_1,...,\mathbf r_k)\Bigg] \, .
	\label{eq:generating_function}
\end{align}
 The derivation of Eq. \ref{eq:generating_function} starting from the statistics of an underlying continuous field is sketched out in more detail in Appendix \ref{sec:derivation}. The shape of the volume $V$ can, in general, be arbitrary. In this paper, we will only consider the volumes to be associated with spheres of radius $r$. As shown in Appendix \ref{sec:derivation}, this is a natural choice for describing statistics of a top-hat smoothed field. In terms of notation, we will switch between $r$ and $V$ throughout the paper, under the implicit assumption $V=4/3\pi r^3$.
 
 The probability of finding a count of $k\in \{0,1,2,...\}$ data points in a volume $V$ can be computed from the generating function by computing various derivatives, 
\begin{equation}
	P_{k|V} = \frac{1}{k!}\Bigg[\bigg(\frac{d}{dz}\bigg)^kP(z|V) \Bigg]_{z=0} \, .
	\label{eq:counts}
\end{equation}
The quantity $P(z|V)|_{z=0}$  or, its mathematical equivalent, $P_{0|V}$ is referred to as the Void Probability Function (VPF) \citep{White1979}, and represents the probability of finding no data points within a volume $V$. Note that the expression for the VPF still contains all $N$-point correlation functions, and \citet{White1979} showed, using a slightly different formalism from that used here, the VPF itself can be considered as the generating function of the full distribution. In the literature, $P_{k|V}$ corresponds directly to the CIC statistics for tracer particles. For scales much larger than the mean inter-particle separation of the tracers, where the mean number of data points per volume, $\langle k_V\rangle \gg 1$, $P_{k|V}$ corresponds to the density PDF of the underlying continuous field \citep[\textit{e.g.}][]{2018MNRAS.481.4588K,2019arXiv191111158U}. Using a very similar formalism, it is possible to write down the generating function for various cumulants of the distribution, which are directly related to the $N$-point connected correlation functions. This is demonstrated in Appendix \ref{sec:cumulants}.

Next, we consider the statistics of volumes which have more than $k$ data points, $P_{>k|V}$, where, once again, $k\in\{0,1,2,...\}$. We will first write down the generating function for this statistic, $C(z|V)$ in terms of the $P(z|V)$. We follow the same definition as Eq.~\ref{eq:generating_function} to write 
\begin{align}
	C(z|V) &= \sum_{k=0}^{\infty} P_{>k|V} z^k = \sum_{k=0}^{\infty} \sum_{m=k+1}^\infty P_{m|V}z^k \nonumber \\
	& = (P_{1|V}+P_{2|V}+...) + (P_{2|V}+P_{3|V}+...)z \nonumber \\
	& \qquad + (P_{3|V}+P_{4|V}+...)z^2 + ... \nonumber \\
	&= -P_{0|V} +(P_{0|V}+P_{1|V}+P_{2|V}+...) + \nonumber \\ 
	& \qquad -(P_{0|V}+P_{1|V})z + (P_{0|V}+P_{1|V}+P_{2|V}+...)z \nonumber \\
	& \qquad -(P_{0|V}+P_{1|V}+P_{2|V})z^2 + (P_{0|V}+P_{1|V}+...)z^2 \nonumber \\
	& \qquad + ... \nonumber \\
	&= -P(z|V)(1+z+z^2+...) + (1+z+z^2+...) \nonumber \\
	&= \frac{1-P(z|V)}{1-z} \, ,
	\label{eq:cdf_generating_function}
\end{align}	
where we have used the fact $\sum_{k=0}^\infty P_{k|V}=1$, and $1/(1-z)=(1+z+z^2+...)$. Therefore, the generating function for the distribution of $P_{>k|V}$ is fully specified by the generating function of $P_{k|V}$. Note that, by definition
\begin{align}
	P_{k|V} =   P_{>k-1|V} - P_{>k|V}\qquad \forall k \geq 1\, .
	\label{eq:cic_cdf}
\end{align}

From Eqs. \ref{eq:generating_function}, \ref{eq:cdf_generating_function}, and \ref{eq:cic_cdf}, it becomes clear there are three equivalent approaches to characterizing the clustering of a set of data points through the generating function: $1$) by measuring all the connected $N$-point correlation functions $\xi^{(N)}$, \textit{i.e.}, the second line in Eq. \ref{eq:generating_function}, $2$) by measuring the distribution of the counts in cell, $P_{k|V}$, \textit{i.e.} the first line in Eq. \ref{eq:generating_function}, and $3$) by measuring the cumulative counts, $P_{>k|V}$, and connecting them to $P_{k|V}$ using Eq. \ref{eq:cic_cdf}. To fully characterize the generating function, each of these statistical measures has to be measured over the full range of scales of interest. In the next subsection, we present the method for efficiently calculating $P_{>k|V}$ concurrently over a large range in $V$ for a set of data points. This is done by exploiting the connection between $P_{>k|V}$ and the $k$-Nearest Neighbor distributions of the data points from a set of volume-filling random points.

\subsection{Efficient calculation of $P_{>k|V}$ in data using $k\nn$-$\cdf$}
\label{sec:computing_nncdf}

\begin{figure}
	\includegraphics[width=\columnwidth]{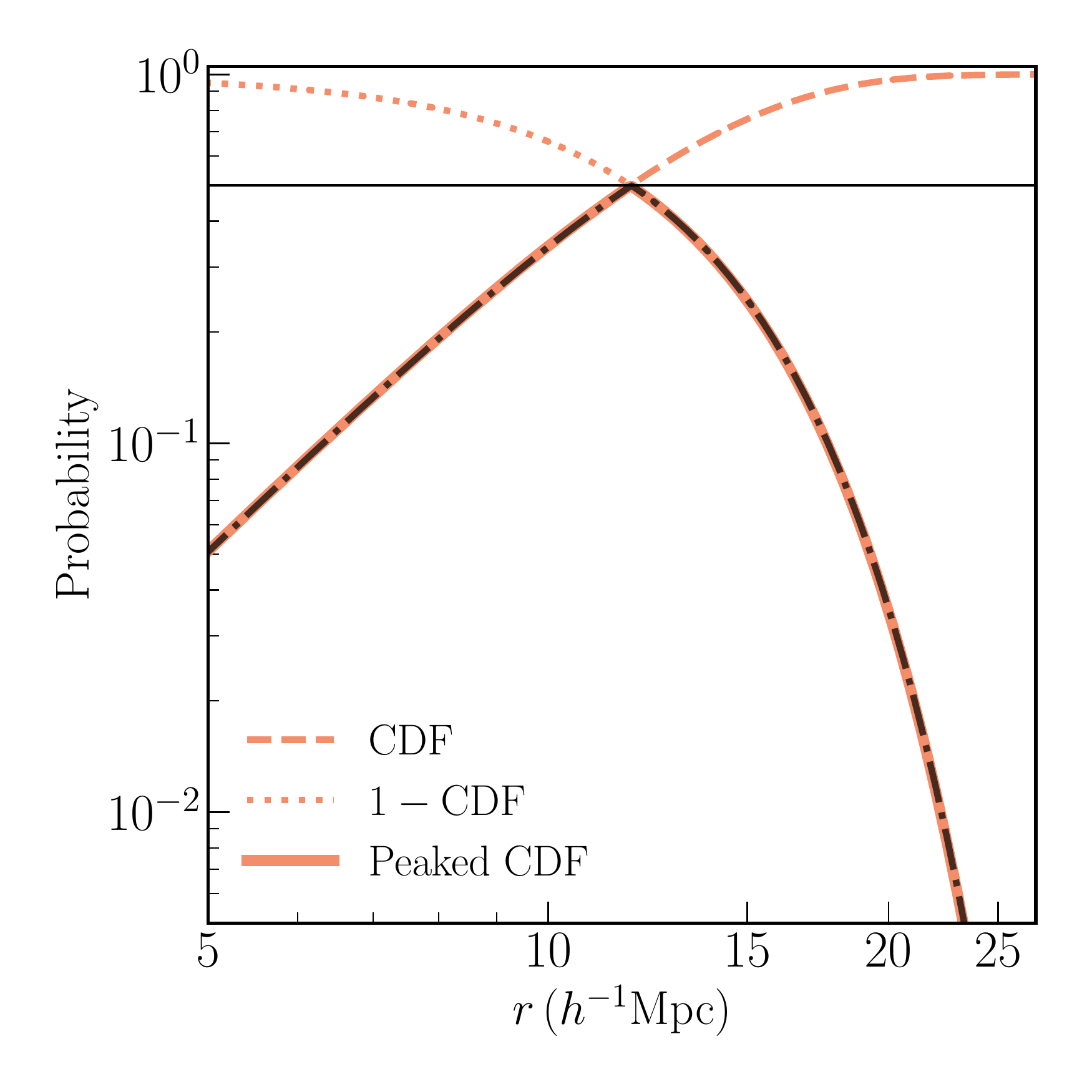}
    \caption{Peaked CDF (described in the text) of the nearest neighbor distribution (as defined in the text) for $10^5$ random points distributed over a $(1 \hgpc)^3$ volume (solid curve). The dot-dashed curve is the analytic prediction for the distribution. The empirical CDF measured from the data is plotted using the dashed curve, while the Void Probability Function (VPF) is plotted using the dotted line.}
    \label{fig:Poisson_nncdf}
\end{figure}

For a set of $N_d$ data points distributed over a total volume $V_{\rm tot}$, we start by generating a volume-filling sample of random points. Typically, the total number of randoms, $N_R$ is chosen to be larger than the number of data points, $N_d$, and using more randoms allows for better characterization of the tails of the distributions discussed below. Once the set of randoms have been generated, for each random, we compute the distance to, say, the nearest data point. This can be done extremely efficiently by constructing a $k$-d tree on the data \citep[e.g.][]{Wald2006} in $N\log N$ operations. Using this tree structure allows to search for the $k$ nearest neighbors in $\log N$ time for each point.
Most scientific software has efficient built--in functions or libraries such as e.g. \textsc {scipy}'s \textsc {cKDTree} implementation, or \textsc{Julia}'s \textsc{NearestNeighbor.jl}\footnote{https://github.com/KristofferC/NearestNeighbors.jl} library. These typically will return an ordered list of the distances to the nearest $k$ neighbors.
Sorting the computed distances for each $k$,  immediately gives the Empirical Cumulative Distribution Function (see \textit{e.g.} \citet{vaart_1998}) of the distribution of distances to the $k$-nearest data point from the set of volume-filling randoms. In the limit of large $N_R$, this converges to the true underlying Cumulative Distribution Function (CDF) of the distance to the $k$-nearest data point from any arbitrary point in the box. Notice that since the randoms are volume-filling, all regions are sampled equally, irrespective of whether the region is overdense or underdense in terms of the data points. This is especially relevant to applications in cosmology, where at late times, the volume is dominated by underdense regions, while the data points are usually concentrated in overdense regions. Measures such as the two-point correlation functions are typically dominated by the overdense regions, and have little information about the large underdense regions~\citep[e.g][]{Oort1983,2013MNRAS.433.1628N}.

\begin{figure*}
	\includegraphics[width=\textwidth]{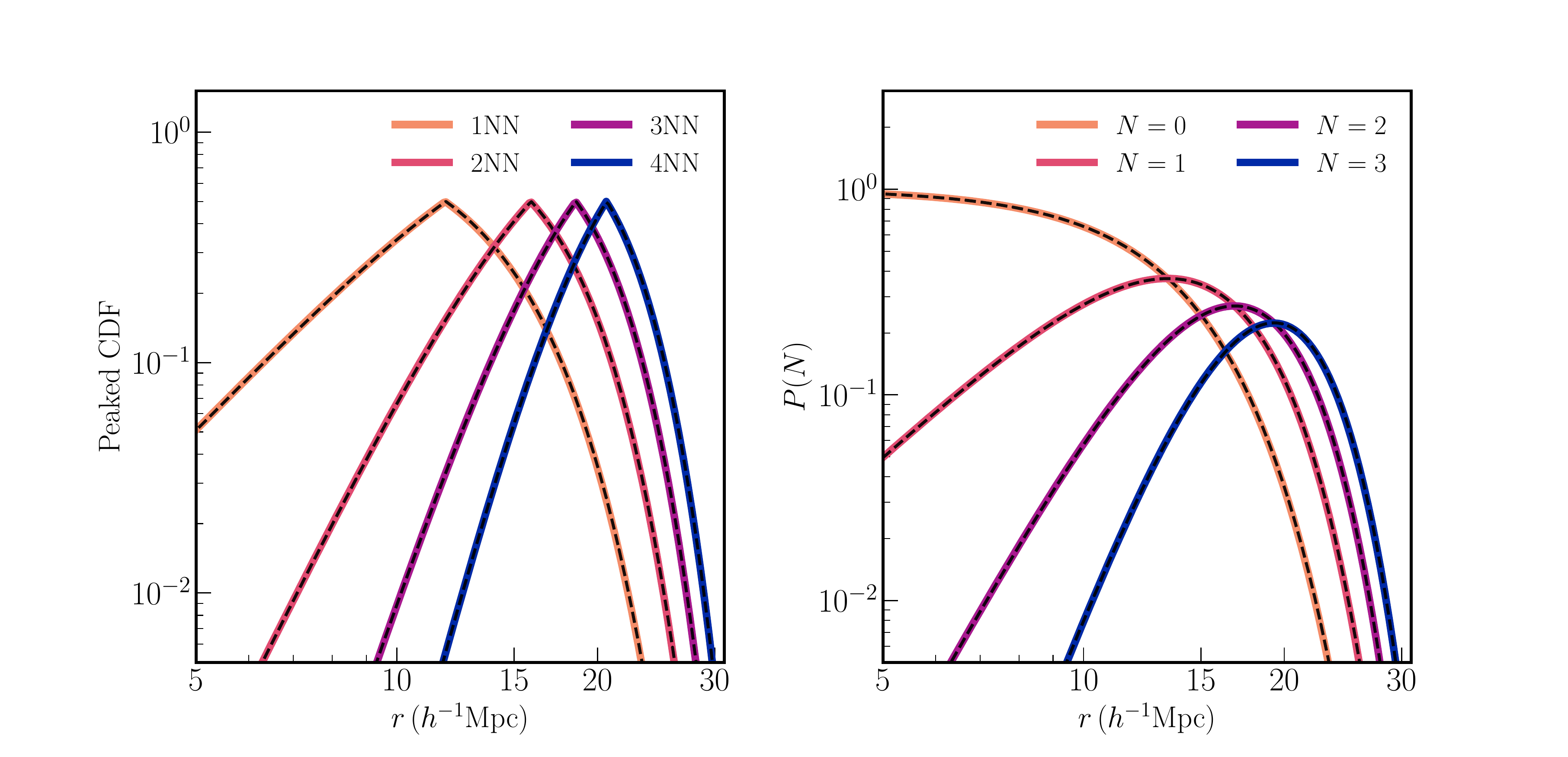}
	\caption{\textit{Left}: Peaked CDF as a function of scale for $1$st (${\rm 1NN}$), $2$nd (${\rm 2NN}$), $3$rd (${\rm 3NN}$), and $4$th (${\rm 4NN}$) nearest neighbor distributions (solid lines) for a set of $10^5$ Poisson distributed data points distributed over a $(1 \hgpc)^3$ volume from a set of volume filling randoms (see text for details). The dashed lines represent the analytic expectations for the distribution. \textit{Right}: Probability of finding $N$ points in a sphere with radius $r$ given  $10^5$ Poisson distributed data points over a $(1 \hgpc)^3$ volume. Solid lines represent the probabilities computed using the CDFs from the left panel, while the dashed lines represent the analytic expectation.}	
    \label{fig:Poisson_nn_cic}
\end{figure*}

To understand the connection between the CDF of the distribution of distances to the nearest data point from a random point in the volume, and the probability of $P_{>k|V}$, consider all possible spheres of volume $V = 4/3 \pi r^3$ \citep{1999ApJ...513..543K}. For a closely related statistic, see \cite{1984ApJ...287L..59R}. These spheres can be centered on any point in the total volume under consideration. The fraction of spheres with $>0$ data points is exactly the fraction of sphere centers for which the distance to the nearest data point is $<r$. The nearest-neighbor CDF at some radius $r$  is the precise measure of the fraction of points for which the nearest data point is at a distance $<r$. Therefore,
\begin{align}
	\label{eq:1nn}
	{\rm CDF}_{\rm 1NN}(r) = P_{>0|V}\big|_{V=4/3\pi r^3} \, .
\end{align}
As mentioned previously, we will switch notations between radius $r$ and the volume $V$ throughout this paper, and we note that none of our results depend on the distinction in notation. Given the connection between nearest neighbor distances and finding a certain number of data points in a given volume, $1 -{\rm CDF}_{\rm 1NN}(r)$ then represents the probability of finding a completely empty volume $V=4/3\pi r^3$. That is, a randomly placed sphere of Volume $V$ is empty with a probability of $1 -{\rm CDF}_{\rm 1NN}(V)$, which is known as the Void Probability Function (VPF, \cite{White1979}).  
Interestingly, this latter interpretation is what is customarily used (e.g. \textsc{Corrfunc} code\footnote{https://github.com/manodeep/Corrfunc} \cite{10.1007/978-981-13-7729-7_1,2020MNRAS.491.3022S}) to measure the VPF using large numbers of randomly placed spheres. This approach is much slower, however, as compared to the kNN--CDF. It also only provides measurements at typically a small number of chosen volumes. So even if one were only interested in computing the VPF, using the nearest neighbor approach discussed here would be advisable since the measurements are desired at multiple volumes. Note that the empirical CDF directly measured from the data contains as many points as random points we chose to cover the full volume. Given the monotonic nature of the CDF, using linear or higher order interpolation between measured points allows one to carry out operations on these CDFs as if they were continuous functions.

We illustrate the shapes of these functions for a Poisson distribution of points in Fig.~\ref{fig:Poisson_nncdf}. While the Empirical CDF is the quantity which is directly computed from the simulations, in our plots we will usually display the Peaked CDF (PCDF) which is defined as
\begin{equation}
	{\rm PCDF(r) = \begin{cases}
		\cdf(r) \qquad &\cdf(r) \leq 0.5 \\
		1 - \cdf(r) \qquad &\cdf(r) >0.5 \, .
	\end{cases}}
\end{equation}
The use of the PCDF allows for better visual representation of both tails of the CDF. This point is illustrated in Fig.~\ref{fig:Poisson_nncdf} for a set of $10^5$ points distributed according to a Poisson distribution over a $(1 \hgpc)^3$ volume. The dashed line represents the Empirical Cumulative Distribution Function measured from the set of particles. The solid line represents the Peaked CDF computed from the same data. The right hand tail of the distribution is difficult to differentiate using the Empirical CDF, since, by its very nature, it asymptotes to 1 smoothly. The Peaked CDF, on the other hand, illustrates clearly the behavior in the tails, especially when comparing to the analytic expectation, plotted using the dot-dashed line. We also plot the VPF ($=1-\cdf_{1\nn}$) using the dotted line for reference. 

Importantly, apart from the distance to the $1$st nearest neighbor data point, the same tree is used to find the distances to the $k$-th nearest data point for each random point in the volume. Once again, these distances can be sorted to produce the Empirical CDF of the $k$-th neighbor distances, and in the limit of large $N_R$, the true underlying CDF of the $k$-th neighbor distances. We can generalize the arguments presented above to connect the probability of finding spheres of volume $V=4/3\pi r^3$ enclosing $> k-1$ data points to the CDF of distances to the $k$-th nearest data point within radius $r$:
\begin{align}
	\label{eq:knn}
	{\rm CDF}_{k{\rm NN}}(r) = P_{>k-1|V}\big|_{V=4/3\pi r^3} \,.
\end{align}
Eq. \ref{eq:cic_cdf} can be recast as
\begin{align}
	P_{k|V} =  {\rm CDF}_{k{\rm NN}}(r) - {\rm CDF}_{(k+1){\rm NN}}(r) \qquad \forall k \geq 1 \, ,
	\label{eq:cic_knn}
\end{align}
\eq{cic_0nn}{P_{0|V} = 1 - \cdf_{1\nn}(r)\, .}
Therefore, computing the $k$ nearest neighbor distributions using the method outlined here is completely equivalent to measuring $P_{>k|V}$. Additionally, this procedure allows us to compute the probabilities over a range of scales, and for multiple values of $k$, in a single operation. To  provide a sense of the time spent on a typical calculation, computing the CDFs for up to $8$ nearest neighbors for $10^6$ randoms and $10^5$ data points distributed in a $(1\hgpc)^3$ volume takes $\sim 20$ seconds on a single core. If the tree search is suitably parallelized, scaling up the number of cores can further reduce the runtime.

We note here that while we have used a set of random points to sample the entire volume, the method outlined above will also work for regularly spaced points, \textit {e.g.} when using points placed on a finely-spaced grid, that is, with grid separations much smaller than the mean inter-particle separation of the data. Once the set of volume filling points are generated, whether using a random procedure, or on the regular grid, the calculation of the distances to the nearest data point, and the computation of the Empirical Cumulative Distribution Function proceeds exactly the same way.

\section{applications to various distributions}
\label{sec:NNCDF_examples}
In this section, we apply the $\nn$-$\cdf$ formalism to tracers following different distributions, and point out various relevant features. We start with the simplest example of a Poisson sampling of a uniform field in Sec. \ref{sec:Poisson}. We then explore the Gaussian distribution in this framework in Sec. \ref{sec:Gaussian}. Finally, in Sec. \ref{sec:LSS}, we apply it to data from cosmological simulations, both simulation particles, and dark matter halos.

\subsection{$k\nn$-$\cdf$ for Poisson sampling of uniform field}
\label{sec:Poisson}

For a sample of Poisson tracers of a uniform field, $\xi^{(1)}=1$, and all higher order correlation functions are $0$. In this case, Eq. \ref{eq:generating_function} simplifies to 
\begin{align}
	P(z|V) = \exp\bigg[\bar n (z-1) V\bigg] \,.
	\label{eq:poisson_generating_function}
\end{align}
As can be anticipated for a pure Poisson process on a uniform field, the expression for the distribution of counts, $P_{k|V}$, in Eq. \ref{eq:counts} becomes
\begin{align}
	P_{k|V} = \frac{1}{k!}\Bigg[\bigg(\frac{d}{dz}\bigg)^kP(z|V) \Bigg]_{z=0}  = \frac{(\bar n V)^k}{k!} \exp (-\bar n V)\, .
	\label{eq:poisson_counts}
\end{align}
The distribution of $P_{>k|V}$ can similarly be worked out by considering the derivatives of $C(z|V)$ from Eq. \ref{eq:cdf_generating_function}:
\begin{align}
	P_{>k|V} &= \frac{1}{k!}\Bigg[\bigg(\frac{d}{dz}\bigg)^kC(z|V) \Bigg]_{z=0} \nonumber \\
	&= \frac{1}{k!}\Bigg[\bigg(\frac{d}{dz}\bigg)^k\Bigg(\frac{1-\exp\big[\bar n(z-1)V\big]}{1-z}\Bigg) \Bigg]_{z=0} \nonumber \\
	& = \frac{1}{k!}\Bigg[\sum_{m=0}^k \frac{k!}{m!(k-m)!}\bigg(\frac{d}{dz}\bigg)^m\bigg(1-\exp\big[\bar n(z-1)V\big]\bigg) \nonumber \\ & \qquad \qquad\bigg(\frac{d}{dz}\bigg)^{k-m} \frac{1}{1-z}\Bigg]_{z=0} \nonumber \\
	&= 1-\sum_{m=0}^k \frac{(\bar n V)^m}{m!} \exp(-\bar n V) \, ,
	\label{eq:poisson_cdf}
\end{align}
where we use the fact that $(d/dz)^m(1/(1-z)) = m!/(1-z)^{m+1}$. The form of $P_{>k|V}$ derived in Eq. \ref{eq:poisson_cdf} can also be anticipated by simply noting that $P_{>k|V} = 1 - P_{<=k|V}$, and using Eq. \ref{eq:poisson_counts}. The form of Eq. \ref{eq:poisson_cdf} is known in the literature as the Cumulative Distribution Function of the Erlang distribution \citep{Peacock_2000}. Since we consider the volumes $V$ to be associated with spheres of radius $r$, all the equations above can also be trivially written in terms of $r$.

Eq. \ref{eq:poisson_cdf} can also be derived in a different way in the language of distances to nearest neighbors ---  by considering the distributions of successive nearest neighbors. Let us consider the CDF of the nearest neighbor distribution: 
\begin{equation}
	{\rm CDF}_{1{\rm NN}}(V) = P_{>0|V} = 1 - \exp(-\bar n V) \, .
\end{equation}
Therefore, at fixed $\bar n$, the PDF of the distribution of distances to, or equivalently the distribution of volumes enclosed within, the nearest data point from a random point is given by 
\eq{nnpdf}{\pdf_{1{\rm NN}}(V) = \frac{d\big(\cdf_{1{\rm NN}}(V)\big)}{dV} = \bar n \exp(-\bar n V) \, .}
Since there are no higher order correlations in the underlying continuous uniform field, the PDF of the distribution of volumes enclosed within the \textit{second} nearest neighbor is a convolution of ${\rm PDF}_{1{\rm NN}}(V)$ with itself:
\eq{convolution_poisson}{\pdf_{2{\rm NN}}(V) &= \int_0^V \pdf_{1{\rm NN}}(V^\prime) \pdf_{1{\rm NN}}(V - V^\prime) d V^\prime \nonumber \\ & = \bar n ^2 V \exp(-\bar n V)\, ,}
and 
\eq{convolution_cdf}{\cdf_{2\nn}(V) &= \int_0^V \pdf_{2\nn}(V^\prime) d V^\prime \nonumber \\ &= 1 - \exp(-\bar n V) - (\bar n V) \exp(-\bar n V)\, .}
Comparing with the last line in Eq. \ref{eq:poisson_cdf}, the expression above is indeed equivalent to $P_{>1|V}$ at fixed $V$. A similar result can be demonstrated for higher $k\nn$ $\cdf$s of the Poisson distribution. For tracers of a Poisson distribution, therefore, computing the distribution of distances to the nearest neighbor data point from any arbitrary point in the volume under consideration is a complete description of the overall distribution of points - distributions of distances to all other neighbors can be generated from the former. Using the connection between nearest neighbor distributions and the probabilities of finding more than $k$ data points in a volume $V$, the above result implies that, for the Poisson distribution, determining $P_{>0|V}$ or the VPF alone allows us to extract maximal information about distribution. This is consistent with the fact that the only variable for a Poisson distribution is the rate ($=\bar n V$ here), and that the full distribution can be generated once the rate is known, from say, the nearest neighbor distribution. Expressed in another way, there is no new information in any of the higher $k$-th neighbor distributions, once the nearest neighbor distribution is known.

We now compare the analytic predictions to the actual measurements from a set of $10^5$ points distributed randomly over a $(1\hgpc)^3$ volume. The results are shown in Fig. \ref{fig:Poisson_nn_cic}, where the first four nearest neighbor distributions are plotted in the left panel. The four lowest counts-in-cell distributions, computed from the nearest neighbor distributions on the left, are plotted in the right hand panel. In both panels, the solid lines represent the measurements, while the dashed line represent the analytic expectations. The measurements in the left panel agree extremely well with the analytic expectations for scales below, and comparable to, the mean inter-particle separation ($\sim 15 \hmpc$). The right panel, where the solid curves have been computed using Eq. \ref{eq:cic_knn}, clearly demonstrates that the nearest neighbor distributions and the counts-in-cells at a given radius are truly equivalent descriptions of the underlying data, and that one can easily be computed once the other is known.

On scales much larger thatn the mean inter-particle separation, the Empirical CDF can deviate from the true underlying CDF for two reasons. The first is due to the limitations on the sampling of the total volume arising from the use a finite number of random particles. In the method outlined in Sec. \ref{sec:computing_nncdf}, the number of randoms determines how well the Empirical CDF is sampled, and finite sampling can lead to discrepancies, especially in tails, where the true CDF is being approximated from a small number of measurements. The second effect is that of sample variance in the data itself. This arises because we evaluate the Empirical CDF on one specific realization of $10^5$ points drawn from a Poisson distribution. In general, we will usually restrict our measurements and analysis to scales which are a few times the mean inter-particle separation of the tracers, and ensure that the measurements are not affected by the lack of sampling in the tail.


\begin{figure*}
	\includegraphics[width=\textwidth]{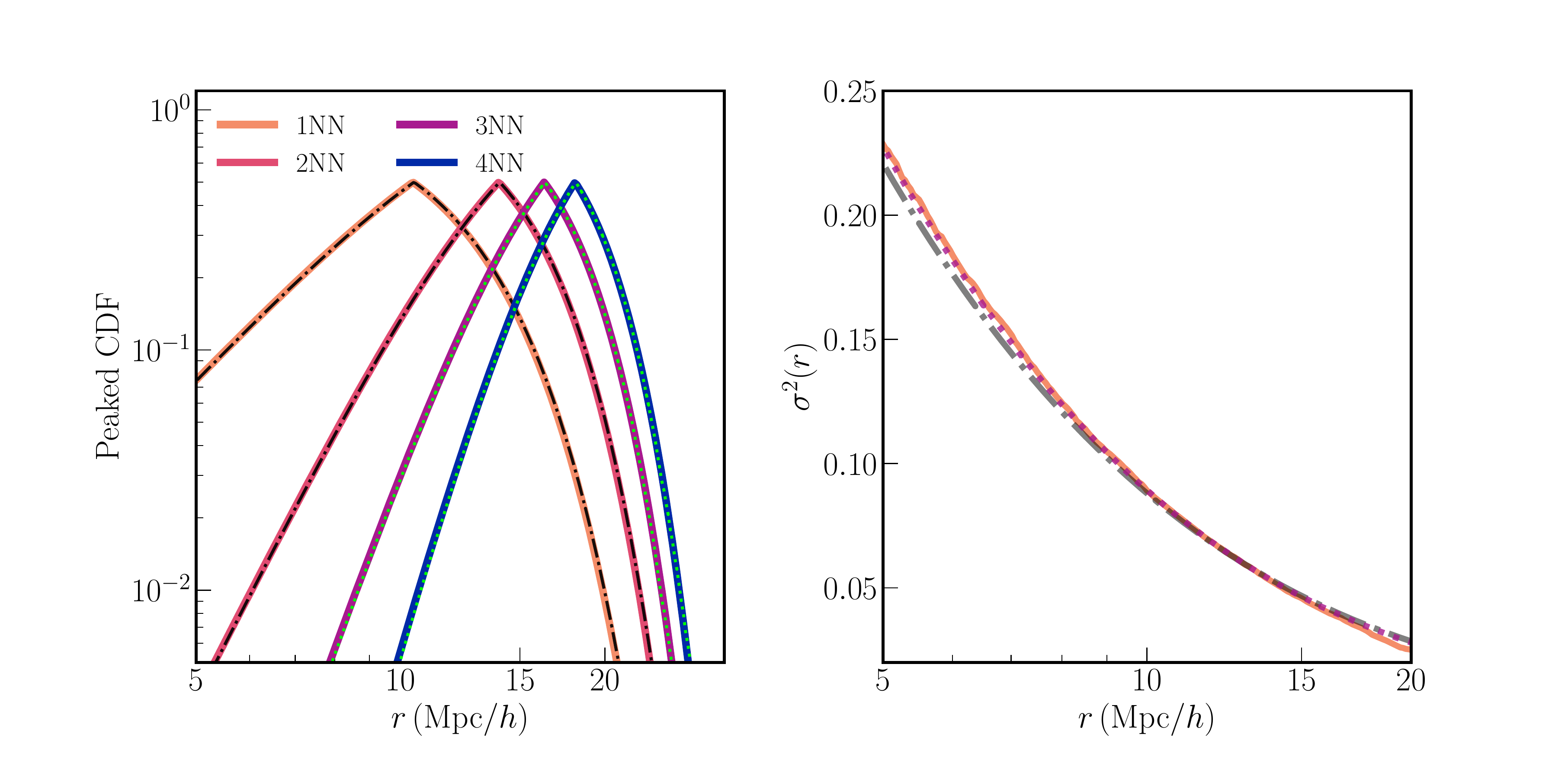}
	\caption{\textit{Left}: Peaked CDF as a function of radius for $1$st (${\rm 1NN}$), $2$nd (${\rm 2NN}$), $3$rd (${\rm 3NN}$), and $4$th (${\rm 4NN}$) nearest neighbor distributions (solid lines) for a set of $10^5$ points tracing an underlying (almost) Gaussian field distributed over a $(1 \hgpc)^3$ volume. The dash-dotted lines for the $1\nn$ and $2\nn$ represent the analytic expectations for distribution given the linear theory variance. The dotted lines represent the predictions for the $3\nn$ and $4\nn$ distributions from the actual measurements of the $1\nn$ and $2\nn$ distributions. \textit{Right}: The solid line represents the variance of the Gaussian field as measured from the $1\nn$ and $2\nn$ CDFs. The dash-dotted line represents the theoretically computed variance from linear perturbation theory. The dotted line is variance measured directly from particles of a higher resolution simulatin (see text for details).}	
    \label{fig:gaussian_nn}
\end{figure*}

\subsection{$k\nn$-$\cdf$ for Gaussian Fields}
\label{sec:Gaussian}

For completely Gaussian continuous fields, the distribution is completely determined by the variance as a function of scale. Therefore the power spectrum, $P(k)$, or the two-point correlation function, $\xi(r)$, are complete descriptions of the underlying field, as well as for any set of tracers of the underlying field. As discussed in Sec. \ref{sec:intro}, these summary statistics have been employed extensively in the study of the CMB, as well as of Large Scale Structure on scales where non-linearities in the density field play a minor role. Therefore, a description of the clustering of a set of tracers of a Gaussian field in the language of $\nn$-$\cdf$s serves as a useful exercise to build up intuition about the connections between the two formalisms before examining a fully non-linear cosmological field. 

For a Gaussian field, we consider the statistics of a set of tracers of the field. These tracers are considered to have been generated by a local Poisson process (see Appendix \ref{sec:derivation} for more details), thereby incorporating the statistics of the underlying field .The generating function for the probability of finding $k$ tracer points in a volume $V$, $P_{k|V}$, can be written as follows:
\eq{gaussian_generating_function}{P(z|V) &= \exp\bigg[\bar n (z-1) V \quad + \nonumber  \\ & \qquad \qquad \frac {\bar n ^2(z-1)^2}{2} \int_V\int_V d^3\mbr_1d^3\mbr_2 \xi^{(2)}(\mbr_1,\mbr_2)\bigg] \nonumber \\ &= \exp \bigg[\bar n (z-1) V + \frac 1 2 \bar n^2 (z-1)^2 V^2 \sigma_V^2\bigg]\, ,}
where $\sigma_V$ is the variance as a function of scale, defined in Eq. \ref{eq:variance}. Notice that the generating function for $P_{k>V}$, given by $C(z|V) = (1-P(z|V))/(1-z)$, therefore, contains only two possible unknowns - the mean number density $\bar n$, and the variance as a function of scale $\sigma_V$. The individual $P_{>k|V}$ distributions can be obtained by taking derivatives of $C(z|V)$. While there is no closed form expression of $P_{>k|V}$ for a general value of $k$, the individual terms are easy to compute, especially for low values of $k$. For example,
\eq{gaussian_counts}{P_{>0|V} &= 1- \exp\bigg[-\bar n V +\frac 1 2 \bar n^2 V^2 \sigma_V^2\bigg] \, , \\ P_{>1|V} &=  P_{>0|V} \nonumber \\ & \qquad - \bigg(\bar n V - \bar n^2 V^2 \sigma_V^2\bigg)\exp\bigg[-\bar n V +\frac 1 2 \bar n^2 V^2 \sigma_V^2\bigg]\, ,}
and so on. Note that just by measuring the first two cumulative distributions, $P_{>0|V}$ and $P_{>1|V}$, one can constrain $\bar n$ and $\sigma_V^2$. Concretely, 
\eq{nv_data}{\bar n V = -2\Bigg(\log\big(1-P_{>0|V}\big)+\frac{1}{2}\frac{P_{>0|V}-P_{>1|V}}{1-P_{>0|V}}\Bigg)\, , }
and 
\eq{variance_data}{\sigma_V^2 = -2\Bigg(\log\big(1-P_{>0|V}\big)+\frac{P_{>0|V}-P_{>1|V}}{1-P_{>0|V}}\Bigg)\bigg/ (\bar n V)^2\, . }
Using the relationship between $P_{>k|V}$ and $\cdf_{(k+1)\nn}(r)$ from Eq. \ref{eq:knn}, Eqs. \ref{eq:nv_data} and \ref{eq:variance_data} can also be expressed in terms of the $1$st and $2$nd nearest neighbor distributions. Once the relevant parameters have been uniquely defined in Eqs. \ref{eq:nv_data} and \ref{eq:variance_data}, all higher $k$ distributions can easily be derived in terms of the measured mean density and variance.

To compare measurements from a set of tracers of a Gaussian field with the theoretical predictions presented above, we consider a very coarse cosmological simulation of $128^3$ particles in a $(1\hgpc)^3$ box at the \textit{Planck} best-fit cosmology run up to $z=3$, when the density field is still roughly Gaussian. We then subsample $1.5\times10^5$ particles from the simulation particles and compute the four nearest neighbor distributions. We use the \textsc{Colossus}\footnote{http://www.benediktdiemer.com/code/colossus/} code to generate the predictions for $\sigma_V$ at this cosmology from linear perturbation theory. The results of the comparison are plotted in the left panel of Fig. \ref{fig:gaussian_nn}. The solid lines represent the measurements of the nearest neighbor distributions from the data, while the dot-dashed line represent the predictions for the first and second nearest neighbor distributions using the theoretical $\sigma^2_V$. Once again, we find good agreement between the measurement and the predictions out to a few times the mean inter-particle separation. The dotted lines in the left panel represent the predictions for the $3$rd and $4$th nearest neighbor distributions, using $\bar n$ and $\sigma_V^2$ measured from the data using Eqs. \ref{eq:nv_data} and \ref{eq:variance_data}. As anticipated from the arguments presented above, measurements of just the first and second nearest neighbor distributions allow us predict all other nearest neighbor distributions to a high degree of accuracy. In the right panel of Fig. \ref{fig:gaussian_nn}, we plot the value of $\sigma_V^2$ that we recover from the measurements of the two nearest neighbor CDFs (solid line), and show that it agrees with the linear theory prediction for the continuous field from \textsc{Colossus} (dot-dashed line), once again on scales comparable to the mean inter-particle separation. We also plot the variance (dotted lines )as computed from the particles of a simulation with $512^3$ particles run to the same redshift. It should be noted that the measurements allow us to correctly infer the variance of the underlying field even on scales smaller than the mean inter-particle separation of the particles used to compute the $k\nn$ distributions ($\sim 12 \hmpc$) --- naively, these scales are expected to be dominated by the Poisson-like $\bar nV$ term in Eq. \ref{eq:gaussian_generating_function}.

We conclude that, for tracers of a Gaussian density field, nearest neighbor distributions beyond the $\second$ nearest neighbor distribution do not add any new statistical information about the underlying field. All higher neighbor distributions can be built up from convolutions of the two nearest neighbor distributions. In fact, these properties of the $\nn$-$\cdf$ for a formally Gaussian field can be turned into a test for the Gaussianity of a given field, or a set of tracers. If the mean density and the variance are computed from the data, then the field is completely Gaussian if and only if the nearest neighbor distributions are consistent with Eq. \ref{eq:gaussian_counts}. Any departures from these expressions can be interpreted as evidence for non-Gaussianity in the field.

In general, therefore, for tracers of a field that is completely characterized by the first $m$ connected $N$-point functions (or cumulants), measurement of only the lowest $m$ nearest neighbor distributions are sufficient to capture the full statistical information of the underlying field. There is no independent information in the higher $\nn$ distributions. It is worth noting that there are certain distributions relevant to cosmological applications, most notably the lognormal PDF, which cannot be uniquely described by its cumulants, or higher order moments (see \cite{2012ApJ...750...28C}, \textit{e.g.}). For such distributions, measuring any finite set of $k\nn$-$\cdf$ is not guaranteed to capture the full statistical description of the underlying field. However, it is worth recalling that while the $k\nn$-$\cdf$ are \textit{related} to the higher order cumulants via the generating function formalism, they are more directly associated with the counts-in-cell (CIC) statistics over a range of scales,as shown in Eq. \ref{eq:cic_knn}. At a fixed radius, $r$, corresponding to volume $V$, a lognormal distribution defines the distribution of $P_{k|V}$, \textit{i.e.}, the distribution of the CIC measurements of data points in spheres of radius $r$. The lognormal distribution, therefore, also defines the distribution of $P_{>k|V}$, in spheres of volume $V$, which are measured directly by the various $k\nn$-$\cdf$ dsitributions. As a result, measuring successively higher $k\nn$ distributions is able to map out the PDF with increasing accuracy, and in principle, capture the full information of even a lognormal distributed field. For a discussion on the information contained in the VPF ($P_0(V)$) for a lognormal field, see \citet{1991MNRAS.248....1C}. It has been pointed out \citep{2018MNRAS.481.4588K} that the density field in $N$-body simulations are not very well-described by the lognormal distribution, and can, in fact, have even more prominent tails on the high density end. In Sections \ref{sec:LSS} and \ref{sec:cosmo_constraints}, we discuss in detail the application of $k\nn$ measurements to density fields and tracers obtained directly from simulations. Even for these fields, it should be kept in mind that we do not know, \textit{a priori}, how many $k\nn$-$\cdf$ need to be measured in order to extract a large fraction of the total information in the field. 

Another aspect to note here is that our choice of downsampling the simulation particles to $1.5\times10^5$ for our measurements was arbitrary, and is not related to the number of particles with which we run the simulation. The choice was guided only the range of scales over which we wish to obtain robust measurements of the $\cdf$. We can also recover the variance of the underlying continuous field for other choices of the mean number density, or equivalently, inter-particle separations. As shown previously, our measurements are most robust on scales comparable to the mean inter-particle separation, so a different choice of the mean number density allows us to accurately measure the variance on a different set of scales. However, the range of scales on which the linear theory variance can be reliably estimated from these measurements is limited on both large and small scales due to numerical and practical considerations. These limitations set the choice of scales displayed in the right panel of Fig. \ref{fig:gaussian_nn}.

First consider the case when the inter-particle separation is much larger than considered here. In principle, this choice should allow us to measure the variance at large scales. However, the shape of the cosmological power spectrum is such that the variance decreases on large scales. If the variance is small on scales which can be well measured for a specific choice of the mean density of tracers, the distributions are dominated by $\bar nV$ term in the exponent on the RHS of Eq. \ref{eq:nv_data}, instead of the $\bar n^2V^2\sigma_V^2$ term. For small enough $\sigma_V^2$, this will be true even when considering scales above the mean inter-particle separation ($\bar n V>1$). Such a scenario makes it numerically difficult to recover the variance of the continuous field using the techniques outlined above. However, it should be emphasized that this is a practical consideration, and not an inherent drawback of the overall method. If the tails of the distribution can be measured extremely accurately - \textit{i.e.} with many more random points than typically used here, it is, in principle, possible to recover the full information about the variance, even at large scales.

At the other end, the problem arises from the fact that in linear theory, $\sigma_V$ increases as we consider smaller scales, until $\sigma_V \sim 1$ on small enough scales. At this point, the field is no longer physical since the implied density PDF has tails which go below $0$. In a simulation, of course, gravitational collapse leads to the generation of higher order correlation functions so that the tracer positions continue to represent a physical density field, as we will explore in more detail in Sec. \ref{sec:LSS}. However, in this case, a measurement of the two nearest neighbor distributions will contain information not just about the mean and the variance, but all higher order moments that are present. This implies that the full clustering statistics can no longer be uniquely determined from just these measurements, and higher $\nn$ distributions contain independent information about the clustering of the underlying field.

\subsection{$k\nn$-$\cdf$ for Large Scale Structure}
\label{sec:LSS}

\begin{figure}
	\includegraphics[width=\columnwidth]{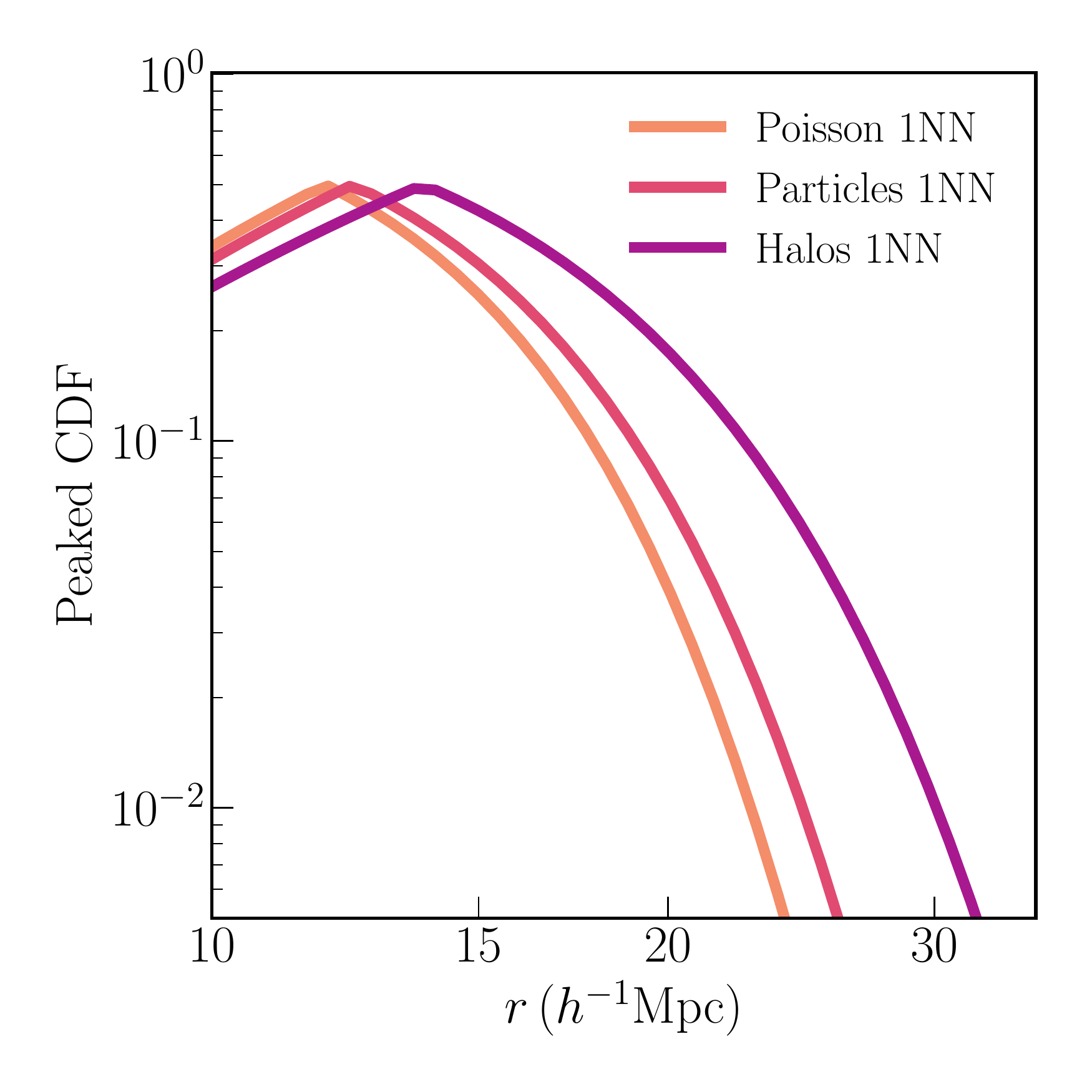}
    \caption{Comparison of the Peaked CDF for nearest neighbor (${\rm 1NN}$) distributions of \textit{a)} a Poisson distribution, \textit{b)} particles from the $z=0$ snapshot of an $N$-body simulation, and \textit{c)} the most massive halos from the same simulation. In each case, $10^5$ points were selected over a $(1 \hgpc)^3$ volume. For the particles, these were randomly selected from all the simulation particles, while for the halos, a cut was made on the $10^5$ most massive halos in the box.}
    \label{fig:LSS_nncdf}
\end{figure}

We now move to analyzing realistic cosmological density fields using the nearest-neighbor formalism that we have set up. At low redshifts, the density field for matter is highly nonlinear, and the distribution of densities on small scales, especially, cannot be approximated by a Gaussian. The clustering of massive virialized dark matter halos, which host the visible galaxies, is usually even more non-Gaussian. In general, the distribution of Dark Matter at low redshifts form a cosmic web - with large empty regions (voids), and high density filaments and knots. In Fig. \ref{fig:LSS_nncdf}, we plot the nearest neighbor distributions for three different sets of points with the same total number of points, $10^5$, spread over a $(1\hgpc)^3$ volume. The  corresponding mean inter-particle separation is $\sim 15\hmpc$. The range of scales represents those where the distributions are best measured for the choice of particle number and the number of randoms that are used to characterize the full volume. The first set of points are distributed randomly over the full volume, \textit{i.e.} following a Poisson distribution. The second set of points are downsampled from the simulation particles in a cosmological simulation with $512^3$ particles at $z=0$. The third set of points are the halo centers of the $10^5$ most massive halos in the simulation at $z=0$. Even though the mean number density is the same, it is clear that the nearest neighbor distributions look quite different. The $\nn$-CDF of the simulation particles, and especially the halo positions, extends to larger scales. This happens due to the presence of large voids in these datasets - this implies that for a large faction of the volume filling set of randoms, the nearest data point is much further away than it would be for a Poisson distribution. We note that while the differences are the most pronounced on larger scales when plotted in terms of the Peaked $\cdf$, the distributions are different even on scales smaller than the mean inter-particle separation.

\begin{figure}
	\includegraphics[width=\columnwidth]{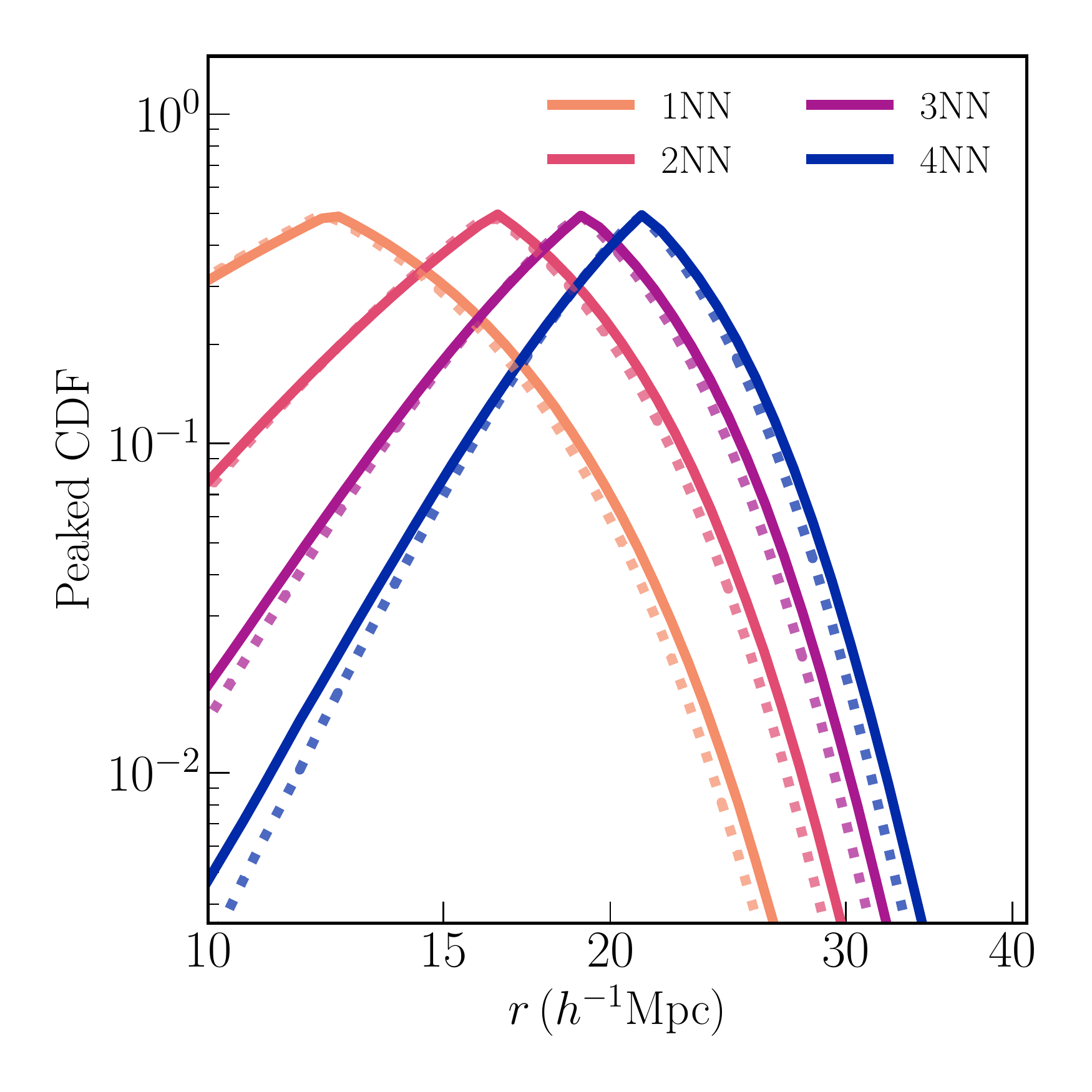}
    \caption{The Peaked CDFs for the first, second, third, and fourth nearest-neighbor distributions for $10^5$ simulation particles in a $(1\hgpc)^3$ volume. The solid lines represent these distributions at $z=0$, while the dotted lines represent the distributions computed at $z=0.5$.}
    \label{fig:LSS_knn}
\end{figure}

It is also useful to plot the change in the nearest neighbor distributions with redshift. Fig. \ref{fig:LSS_knn} shows the first four $\nn$ distributions for $10^5$ simulation particles at z=0 (solid lines) and $z=0.5$ (dotted line). Gravitational evolution of large scale structure drives overdense regions to become more overdense, while underdense regions become more underdense and expand in size. Both these effects can be seen in Fig. \ref{fig:LSS_knn} - on large scales, the distributions extend out further at $z=0$ compared to $z=0.5$ as a result of the voids becoming larger. On small scales, the distributions, especially for the third and fourth nearest neighbors have larger values at a fixed scale for $z=0$ compared to $z=0.5$, as a result of the collapse of the overdense regions.

\subsubsection{Breaking the $b-\sigma_8$ degeneracy}
\label{sec:b_s8}

\begin{figure*}
	\includegraphics[width=0.98\textwidth]{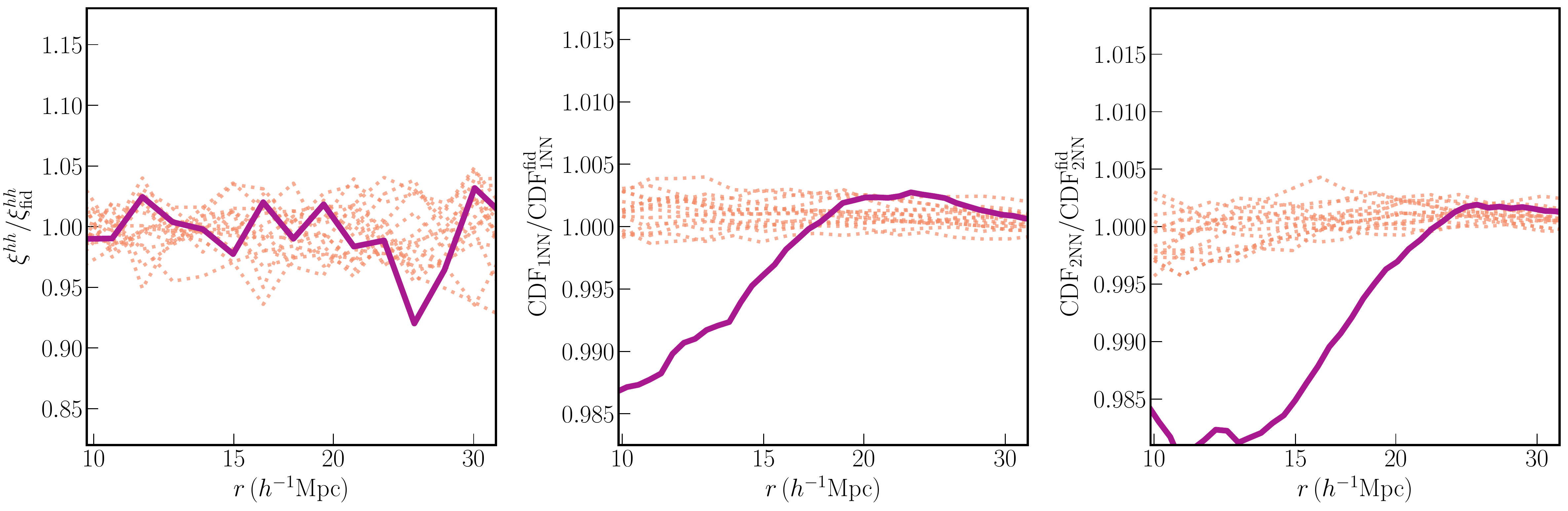}
	\caption{\textit{Left}: The darker line represents the ratio of correlation function of the $10^5$ most massive  halos in a $(1 \hgpc)^3$ box at redshifts $z=0$, and $z=0.5$. The lighter shaded lines represent the ratio of the correlation functions at $z=0$ for $15$ different realizations of the same cosmology, divided by the mean correlation function at that cosmology. \textit{Center}: The darker line represents the ratio of the nearest neighbor CDF of the $10^5$ most massive  halos in a $(1 \hgpc)^3$ box at redshifts $z=0$, and $z=0.5$. The lighter shaded lines represent the ratio of the nearest neighbor CDFs at $z=0$ for $15$ different realizations of the same cosmology, divided by the mean nearest neighbor CDF at that cosmology. \textit{Right}: Same measurements as the center panel, except with second nearest neighbor distances instead of  the first. Even though the correlation function of the two samples at different redshifts are almost indistinguishable within sample variance uncertainties, the $\nn$ $\cdf$s are clearly separated.}
	\label{fig:bias_sigma8}
\end{figure*}

In the analysis of the clustering of halos using the two-point correlation function, there is a known degeneracy between the bias of the halos being considered and the amplitude of clustering of the underlying matter field, sometimes represented by the width of the density PDF at $8 {\rm Mpc}$, $\sigma_8$. This is related to the fact that halos form at the peaks of the initial Gaussian random field \citep{1986ApJ...304...15B}. In other words, two halo populations can have the same two-point clustering signal, even when they are produced by very different underlying matter fields, just by appropriately choosing the bias of each sample. Since the bias is primarily dependent on the halo mass, this is equivalent to making choices about the mass cut for the samples. As a result, it is difficult to individually constrain the value of the bias $b$, and the amplitude of the clustering of the underlying field $\sigma_8$, given a two-point measurement of a sample of halos.

This degeneracy is illustrated in the left panel of Fig. \ref{fig:bias_sigma8}. The dark solid line represents the ratio of the $\xi(r)$ at $z=0.5$ to $\xi(r)$ at $z=0$ for the $10^5$ most massive halos for a single realization at a fixed cosmology. The halos are identified at the different redshifts separately, and so the actual objects that make the cut at different redshifts need not be the same. The measurements are made using halos identified in one of the simulations from the \textsc{Qujiote} suite, discussed in further detail in Sec. \ref{sec:fisher}. The dotted lines represent the ratio of $\xi(r)$ for $15$ different realizations at $z=0$ for the same cosmology, once again using simulations from the \textsc{Quijote} suite, compared to the mean $\xi(r)$ over the $15$ realizations. These curves serve as a measure of the sample variance for this measurement at $z=0$. We can conclude that the two-point functions of the $10^5$ most massive halos in the box at $z=0$ and $z=0.5$ cannot be distinguished over these range of scales to within sample variance, even though the amplitude of clustering of the underlying matter field is quite different at the two redshifts.

As we have shown in Sec. \ref{sec:formalism}, the nearest neighbor distributions are sensitive not just to the two-point correlation functions, but all higher order correlations. This extra information can be used to break the degeneracy outlined above, as has already been demonstrated, \textit{e.g.} in \cite{2005MNRAS.362.1363P} and \cite{2006PhRvD..74b3522S}, by including the bispectrum or three point correlation function in the analysis. We illustrate this in the middle and right panels of Fig. \ref{fig:bias_sigma8}. In the middle panel, the dark solid line represents the ratio of the nearest neighbor CDF of the $10^5$ most massive halos at $z=0.5$ to the nearest neighbor CDF of the $10^5$ most massive halos at $z=0$, for a single realization at a fixed cosmology. The lighter dotted lines represent the ratio of the nearest neighbor CDFs of the $10^5$ most massive halos at $z=0$ for $15$ different realizations of the same cosmology to the mean distribution. Once again the dotted curves serve as a visual representation of the sample variance for the measurement. The right panel repeats the same calculation for the \textit{second} nearest neighbor distribution. It easy to distinguish between the distributions at the two redshifts, either by considering the nearest neighbor distribution or the second nearest neighbor distribution, even though the clustering of the two samples cannot be distinguished by measurements of the two-point correlation function. This result can also be interpreted as proof that at $z=0.5$ and $z=0$, the clustering is quite non-Gaussian on the scales considered. If the halo field was completely Gaussian, Eq. \ref{eq:gaussian_generating_function} would imply that two fields - one at $z=0.5$ and at $z=0$ - with the same variance, as is the case here, would have exactly the same $k\nn$ statistics. The fact that the two fields have different $k\nn$ statistics, despite having the same variance, implies the non-Gaussianity of the clustering.

We note that the fact that the same number density cut - $10^5$ in a $(1\hgpc)^3$ volume - yields the same two-point correlation measurement is a coincidence. The results above would hold even if the number density cuts were different. However, it is also important to note that for the $\nn$-$\cdf$ analysis, care should be taken to use the same number of data points. The distributions depend sensitively on the mean number density, and a change in the mean number density can be misinterpreted as a change in the clustering signal. It is easy to ensure the two datasets have the same number of points by randomly downsampling the larger dataset. We will use this strategy of keeping the number density fixed when using the $\nn$ distributions to analyze different cosmologies and obtain constraints in Sec. \ref{sec:cosmo_constraints}.

\section{Cosmological parameter constraints}
\label{sec:cosmo_constraints}

In this section, we explore the degree to which constraints on various cosmological parameters improve when the same simulation datasets, using the same scale cuts, are analyzed using the nearest neighbor framework developed above, compared to the traditional two-point analysis. We focus on the following range of scales, $10\hmpc$ to $40\hmpc$, \textit{i.e.}, scales smaller than those where the linear, Gaussian approximation is valid, but larger than the typical sizes of even the largest virialized structures in the universe. We avoid using smaller scales, where the gains from using the nearest neighbor statistics are potentially even larger, in the analysis to avoid any possible systematics related to the resolution of the simulations used in the analysis.

\subsection{Fisher formalism}
\label{sec:fisher}

The Fisher matrix formalism has been widely used to estimate the constraints on cosmological parameters given a set of summary statistics, from which the relevant ``data vector'' is constructed, and the expected error bars on the measurement of these summary statistics. For a cosmological survey, the error bars depend on specifications such as the sky area covered by the survey, the depth, and the number density of tracers. When the summary statistics under consideration are two-point correlation functions, or power spectra, the error bars are relatively easy to compute once the survey specifications are known. Therefore, the Fisher formalism can be used to estimate constraints on cosmological parameters around some fiducial cosmology, even before the survey starts collecting data. 

For summary statistics other the two-point correlation function, it is often not possible to analytically compute the error bars, or more generally, the covariance matrix between various entries in the data vector. The Fisher framework has been applied in such situations to estimate the parameter constraints from mock datasets. These datasets are often generated from cosmological simulations, where these non-trivial summary statistics can computed directly from the simulation outputs. This is the spirit in which we use the Fisher formalism in this work. 

Formally, the elements of the Fisher matrix ($\mathbf F$) is defined as 
\eq{Fisher}{\mathbf F_{\alpha\beta} = \sum_{i,j}\frac{\partial D_i}{\partial p_\alpha} \bigg[\mathbf C^{-1}\bigg]_{ij} \frac{\partial D_j}{\partial p_\beta}\, ,}
where $D_i$ are the entries of the data vector, $p_\alpha$ represent various cosmological parameters, and $\mathbf C$ is the covariance matrix for the data vector, evaluated at some fiducial cosmology. The Fisher matrix can then be inverted to determine the constraints on individual parameters, while marginalizing over the uncertainties in all other parameters. as well as the covariances between different parameters. In particular,
\eq{fisher_constraints}{\sigma_\alpha = \sqrt{\left(\mathbf F^{-1}\right)_{\alpha\alpha}}\, ,}
where $\sigma_\alpha$ represents the $1$-$\sigma$ constraint on parameter $\alpha$.

To construct both the derivatives of the data vector with respect to the cosmological parameters, as well as the covariance matrix, we use the \textsc{Quijote}\footnote{https://github.com/franciscovillaescusa/Quijote-simulations} simulations \citep{2019arXiv190905273V}. These $N$-body simulations were run on $(1\hgpc)^3$ volumes with $512^3$ CDM particles in cosmologies with no massive neutrinos, and with $512^3$ CDM and $512^3$ neutrino particles in cosmologies with massive neutrinos. The mean inter-particle separation for the particles is therefore $\sim 2\hmpc$, and to be conservative in our analysis, we only use measurements above $10\hmpc$ in our analysis. The cosmological parameters that are included in the analysis are $\left\{\Omega_m, \Omega_b, \sigma_8, n_s, h, M_\nu, w\right\}$. The \textsc{Quijote} simulations have been run in a way that the derivatives with respect to each of these parameters can be easily computed around a fiducial cosmology. These simulations have already been used to estimate the information content of various non-trivial statistics of the cosmological field \citep{2020JCAP...03..040H,2019arXiv191111158U}.In general, this is done by running simulations with one parameter larger (and smaller) than the fiducial cosmology, while all other parameters are held at their fiducial value. Special care has to be taken to compute the derivatives with respect to the total neutrino mass $M_\nu$, and this is discussed in detail in \citet{2019arXiv190905273V}.

The data vector for the analysis is constructed from the measurement of the $k\nn$-$\cdf$ for $k=\{1,2,4,8\}$. Each distribution has a different functional dependence on all the $n$-point functions present in the data. It also worth reiterating that we do not \textit{a priori} know which $n$-point functions are relevant for the clustering. Using multiple values of $k$, spread over a relatively wide range, as with our particular choice, ensures that the data vector has fewer degeneracies. We have checked that using slightly different combinations of $k$ do not have a major effect on the cosmological constraints. We discuss this further below.
We use $10^5$ data points (simulation particles or halos), and $10^6$ volume-filling random points to generate the Empirical CDF as outlined in Sec. \ref{sec:computing_nncdf}. Each CDF is interpolated to determine its value at $16$ logarithmically spaced values of $r$. Since the analysis is focused on scales from $10\hmpc$ to $40\hmpc$, the values of $r$ for all the CDFs lie within this range. However, to ensure that we do not go too deep into the tails where the finite number of random points starts to affect the ability to accurately measure the distributions, we impose stricter scale cuts for each $\nn$-$\cdf$. For each $k$, we determine the range of scales for which the Erlang CDF distribution (see Eq. \ref{eq:poisson_cdf}) for that $k$, and the same mean number density, lies between $0.005$ and $0.995$. As can be seen from Fig. \ref{fig:LSS_nncdf}, using the Erlang distribution is a conservative choice for the analysis of both simulation particles, and halos, since the tails of the latter distributions extend out further than the reference Poisson distribution. We then choose the $16$ logarithmically spaced $r$ for each $k$ over the range of scales that are allowed after taking into account both the cuts of $10\hmpc$ to $40\hmpc$, and the cuts implied by the Erlang distribution tails. We then append the $16$ measurements for each $k$ into a single data vector with $64$ entries. In most of analysis, we combine measurements from $z=0$ and $z=0.5$, by once again combining the $64$ entry data vector computed for each redshift into a single data vector with $128$ entries. 

Once the data vector is defined and computed on every simulation, the derivative term for each parameter in Eq. \ref{eq:Fisher} is computed by averaging over $100$ realizations at the relevant cosmologies. This helps reduce the noise from both sample variance, and the fact that these derivatives are computed numerically. For Fisher matrix analysis, care needs to be taken that the derivatives are smooth, since numerical noise can lead to spurious features in derivatives which are then interpreted as artificially tight constraints. We inspect the derivatives to ensure that no such pathological features exist over the range of scales we use in the analysis. Some of the derivatives are discussed further in Sec. \ref{sec:cosmo_constraints_particles}.

The covariance matrix is computed from $1000$ realizations at the fiducial cosmology. The entries of the raw covariance matrix are given by 
\eq{covmat_entries}{\mathbf C^\prime_{ij} = \Bigg\langle \bigg(D_i - \langle D_i\rangle\bigg) \bigg(D_i - \langle D_i\rangle\bigg)\Bigg \rangle \, ,}
where the labels $i,j$ represent various rows in the data vector and $\langle ...\rangle$ represents an average over realizations. Note that we set the off-diagonal terms of data vector entries corresponding to two different redshifts to $0$. This is done to avoid spurious covariances arising from the use of the same realizations of the cosmology at $z=0$ and $z=0.5$. To compute the correct inverse covariance matrix that goes into the Fisher analysis, we use the Hartlap correction factor \citep{2007A&A...464..399H}:
\eq{Hartlap}{\mathbf C^{-1} =  \frac{n-p-2}{n-1}\left( \mathbf C^\prime\right)^{-1}\, ,}
where $p$ is total number of entries in the data vector, and $n$ is the number of simulation realizations used in the evaluation of $\mathbf C^\prime$. Notice that this constrains the allowed length of the data vector, given the number of simulations available. For accurate error forecasts, the Hartlap factor should be as close to unity as possible. Since our aim is to only demonstrate the gain in constraining power when using $k\nn$ analysis over $\xi(r)$ analysis for the same analysis choices, and not the actual values of the constraints, we use the native simulation volume of $(1\hgpc)^3$ throughout our analysis. Current and future cosmological surveys typically have much larger volumes, and consequently can lead to much tighter absolute constraints than the ones presented here, even after accounting for possible systematics arising in the data. While inverting the covariance matrix, we  check that the condition number is within acceptable limits, for each of the analyses presented below. We also explicitly check that the distribution of deviations of the data vector around the mean are roughly Gaussian for different entries in the data vector, implying that the Fisher error estimates should be roughly valid.

We use exactly the same framework to compute the constraints on cosmological parameters from the two-point correlation function over the range $10\hmpc$ to $40\hmpc$. At each redshift, we compute $\xi(r)$ in $30$ logarithmically spaced bins. We then combine the data vectors at $z=0$ and $z=0.5$ into a single data vector. The rest of the analysis proceeds the same way as outlined above. In principle, the Fisher analysis on the two-point function could also have been carried out using analytic estimates of the covariance matrix. However, using the same pipeline for both observables helps keep both analyses consistent in terms of numerics, while also serving as a systematics check on the $k\nn$-$\cdf$ results.

\begin{figure}
	\includegraphics[width=\columnwidth]{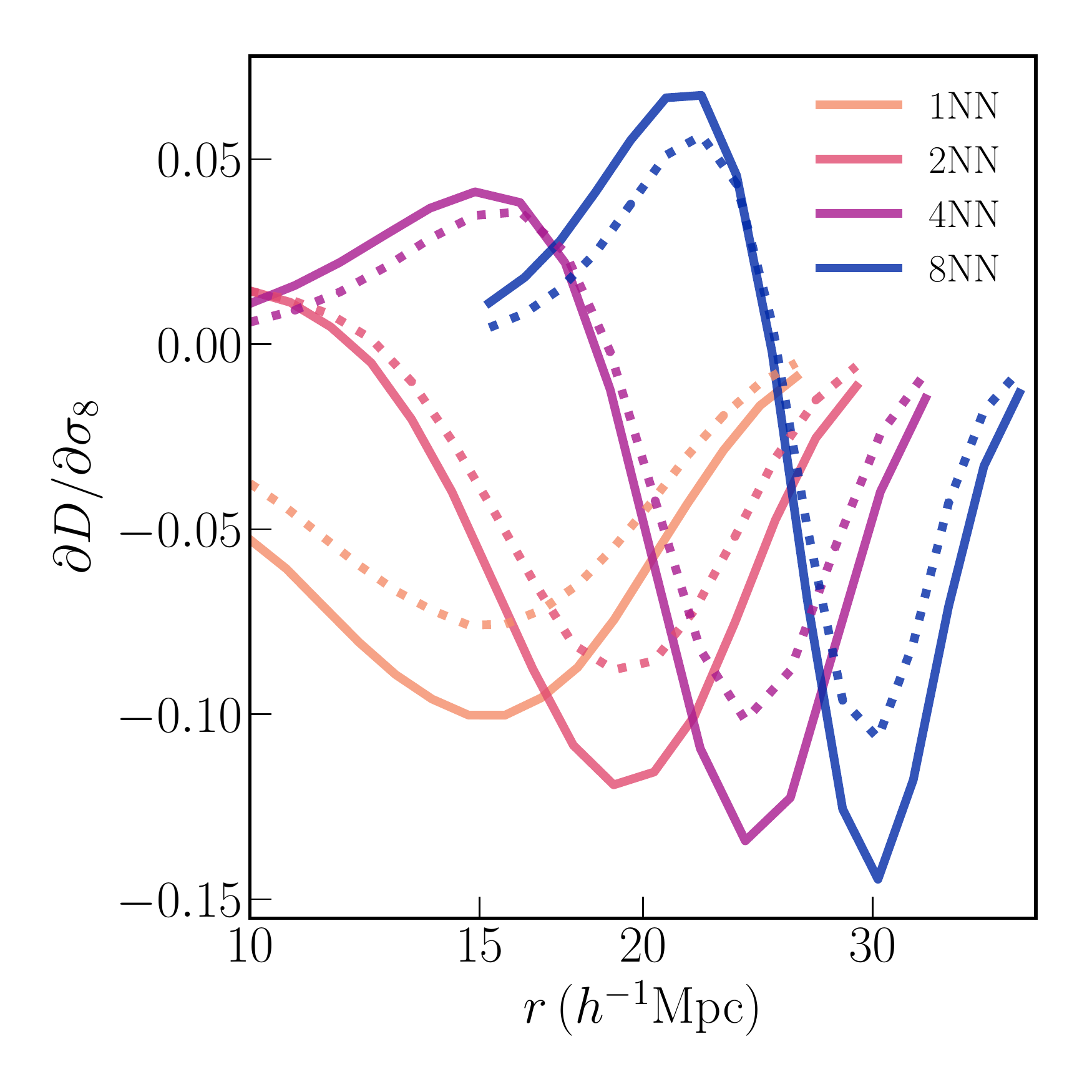}
    \caption{The derivative of the data vector with respect to the cosmological parameter $\sigma_8$. The different colored curves represent the portions of the data vectors coming from $k=\{1,2,4,8\}$ nearest neighbor CDF distributions. The solid lines represent the derivative at $z=0$, while the dotted lines represent the derivative at $z=0.5$.}
    \label{fig:s8_derivative}
\end{figure}

\begin{figure}
	\includegraphics[width=\columnwidth]{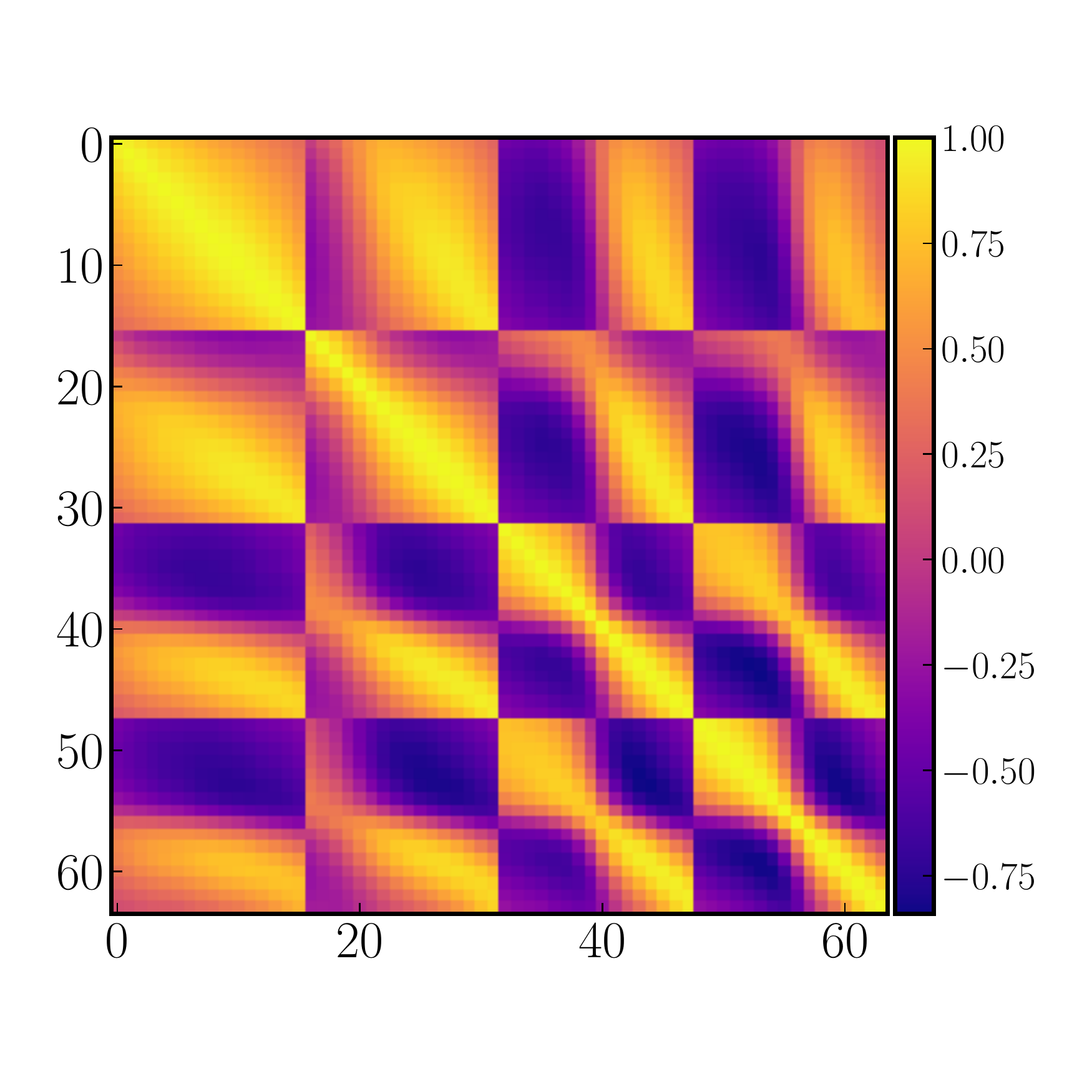}
    \caption{The correlation matrix, as defined in the text, for data vector entries corresponding to measurements at $z=0$. The co-ordinates on the $x$ and $y$ axis correspond to entries in the data vector. Each sub-block corresponds to entries from $k=\{1,2,4,8\}$ nearest neighbor measurements.}
    \label{fig:correlation_matrix}
\end{figure}

\subsection{Constraints from Dark Matter density field}
\label{sec:cosmo_constraints_particles}

\begin{figure}
	\includegraphics[width=\columnwidth]{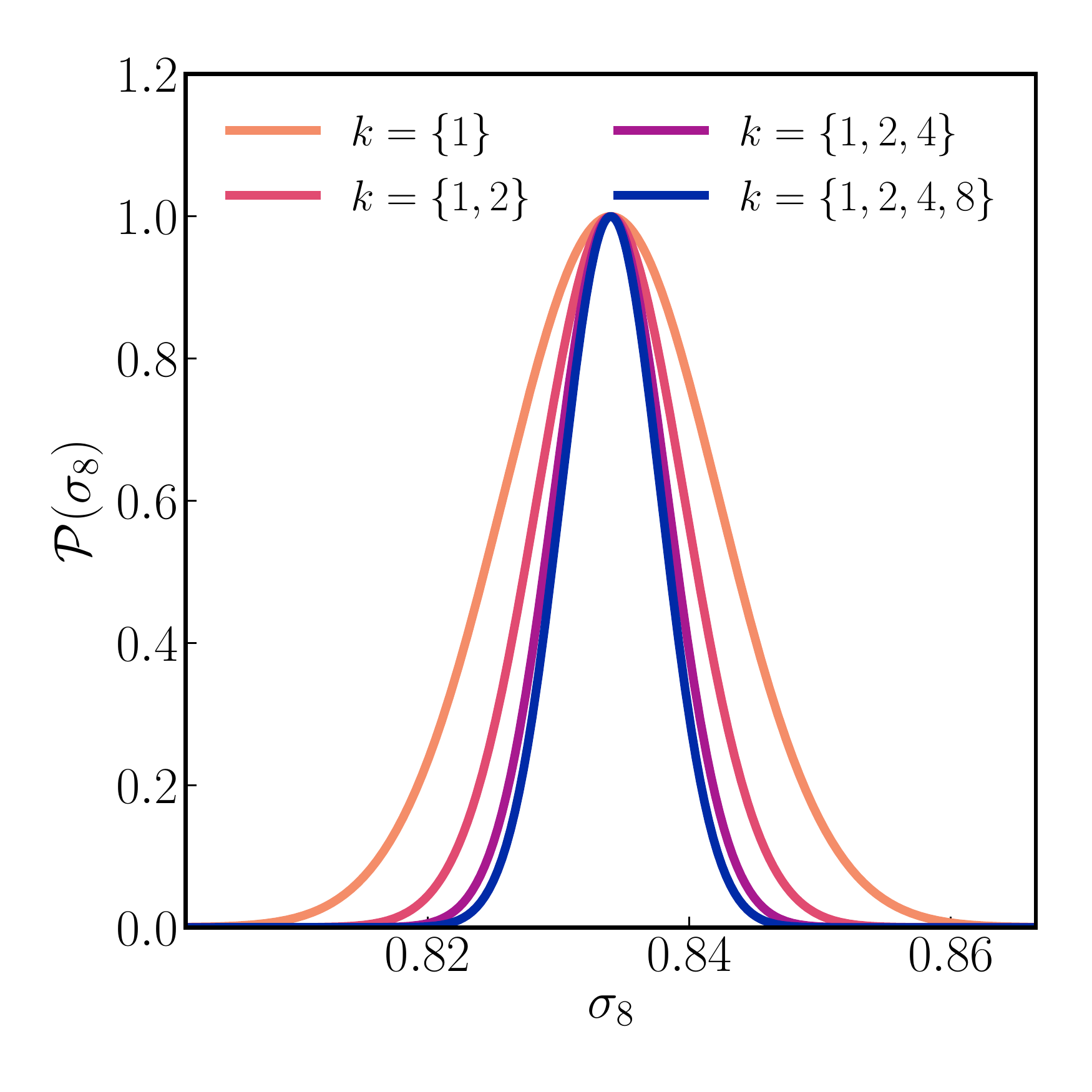}
    \caption{The posterior distribution for $\sigma_8$, marginalized over all other parameters. The different colors represent different $k\nn$ combinations from which the constraint was obtained. The constraints improve as more nearest neighbor distributions are added, but the gain saturates by the time we add all four CDFs that are computed from the data. }
    \label{fig:s8_posterior}
\end{figure}

\begin{figure*}
	\includegraphics[width=0.9\textwidth]{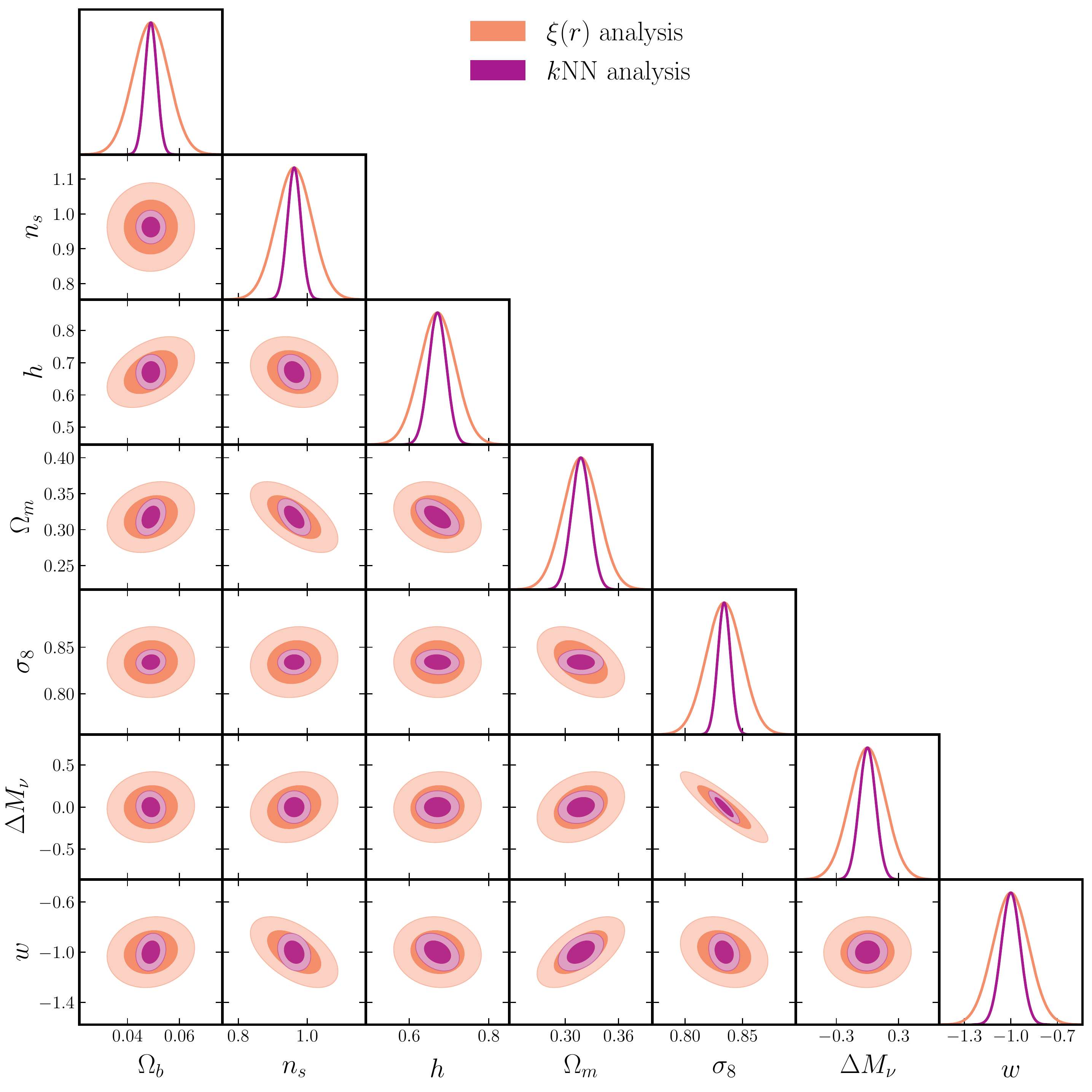}
	\caption{Constraints on the cosmological parameters derived from the Fisher analysis on simulation particles, combining information from $z=0$ and $z=0.5$, using scales in the range $10\hmpc$ to $40\hmpc$. The constraints from the nearest neighbor analysis (using the same set of scales) are tighter by more than a factor of $2$ for most of the parameters. The values of the constraints on individual parameters are listed in Table \ref{tab:part_constraints}.}
	\label{fig:part_constraints}
\end{figure*}


In this section, we present the constraints on the cosmological parameters when considering the $3$-dimensional Dark Matter density field, where the simulation particles are tracers of the field. For the nearest neighbor analysis, we downsample the simulation particles to $10^5$ before computing the different distributions. As shown in Sec. \ref{sec:Gaussian}, the downsampling determines the range of scales over which we can accurately measure the features of the underlying continuous field. The specific choice we make here is to ensure that the measurements are robust in the range of scales that enter the Fisher analysis, i.e. $10\hmpc$ to $40\hmpc$. To reduce the sampling variance caused by downsampling, for every cosmology, we create $16$ distinct downsamplings of $10^5$ particles each from the original $512^3$ particles in the simulation, and then compute the Empirical CDF by combining the nearest neighbor measurements from individual downsampled datasets. For the correlation functions, we use the \textsc{Corrfunc} code\footnote{https://github.com/manodeep/Corrfunc} \citep{10.1007/978-981-13-7729-7_1,2020MNRAS.491.3022S} to compute $\xi(r)$ using all $512^3$ particles in the simulation. Once the data vector is defined in this way, we compute the covariance matrix and the derivatives of the data vector with respect to the cosmological parameters. Note that we use only the CDM particles, for both the nearest neighbor analysis, and the $\xi(r)$ analysis, even for cosmologies with massive neutrinos. In massive neutrino cosmologies, the CDM density field, traced by the CDM particles, and the total matter field, which determines quantities relevant for gravitational lensing, for example, are different. The total matter field includes the contribution from the clustering of neutrinos, and is known to be more constraining on the total neutrino mass $M_\nu$. For this work, we only present the constraints on the cosmological parameters, including the neutrino mass, from the CDM field only. We will explore the possible stronger constraints when using the total matter field in a future work.

In Fig. \ref{fig:s8_derivative}, we plot the derivative of the data vector with respect to $\sigma_8$. The different colors represent the parts of the data vector that come from different $k\nn$ distributions. The solid lines represent the part of the data vector computed from particles at $z=0$, and the dashed lines represent the parts of the data vector computed from particles at $z=0.5$. The sign of the derivative for the part of the data vector coming from nearest neighbor distribution can be understood in the following way --- when the amplitude of clustering at $8{\rm Mpc}$, $\sigma_8$ is higher, there are more empty void-like regions on larger scales. In other words, on these scales, the probability of finding $0$ particles (VPF) in a sphere of radius $r$ is higher when $\sigma_8$ is higher. Since $\cdf_{1\nn}(r)=P_{>0|r} = 1-{\rm VPF}(r)$, a higher $\sigma_8$ implies a lower value for the data vector entry at these scales if all other cosmological parameters are held fixed. This leads to the negative sign of the derivative seen in Fig. \ref{fig:s8_derivative}. The derivative becomes larger in magnitude as the redshift decreases, as is expected from the growth of structure with time. We check that all the other derivatives are smooth, and do not suffer from numerical artifacts on scales that enter the analysis.

Next, we explore the structure of the correlation matrix, $\tilde {\mathbf C}$, of the $k\nn$ measurements. The individual elements of $ \tilde {\mathbf C}$ are defined as
\eq{eq:correlation_matrix}{\tilde{\mathbf C}_{ij} = \frac{\mathbf C_{ij}}{\sqrt{\mathbf C_{ii}\mathbf C_{jj}}}\, .}
We plot the correlation matrix for measurements at $z=0$ are plotted in Fig. \ref{fig:correlation_matrix}. The measurements at $z=0.5$  produce qualitatively similar results. The $x$ and $y$ coordinates of Fig. \ref{fig:correlation_matrix} correspond to entries in the data vector, while each of the visible sub-blocks corresponds to measurements of $k=\{1,2,4,8\}$ nearest neighbor distributions. Within each sub-block, we find significant correlations between various radial scales - this is not surprising since the entries are computed from a CDF. Variations in measurements at small scales propagate to large scales through the use of the cumulative value. There are also significant correlations between measurements at different $k$. Once again, this is to be expected, given that we measure $P_{>k|V}$. For any $k>1$, this measurement includes contributions from all $P_{>j|V}$ measurements for $j<k$.

Before moving to the combined constraints on all cosmological parameters, we first show how the constraints on one of the parameters, $\sigma_8$, changes as we add information from different $k\nn$ distributions at $z=0$. For this example, we first use a data vector constructed only from the nearest neighbor CDF at z=0, and calculate the constraints on $\sigma_8$ from the Fisher analysis on this data vector. Next, we use the measurements from $k=\{1,2\}$ nearest neighbors, and repeat the calculation. We repeat this until we use all four $k=\{1,2,4,8\}$ nearest neighbor CDFs in our analysis. The posterior distribution for $\sigma_8$ from each of these calculations is plotted in Fig. \ref{fig:s8_posterior}. Note that this posterior is marginalized over all other cosmological parameters even though they are not shown here. We find that the constraints improve as more $\nn$ distributions are added to the data vector. However, the gain from adding new $\nn$ distributions diminishes by the time we use all four of the computed CDFs, $k=\{1,2,4,8\}$. A similar trend is observed for other cosmological parameters as well. We conclude that up to the lowest scales in the analysis, the choice of $k=\{1,2,4,8\}$ in our nearest neighbor analysis is sufficient to extract most of the information on the cosmological parameters. If smaller scales are to be included in the analysis, higher $k$ neighbors may have to be considered to ensure maximal constraints on the parameters down to those scales.

The joint constraints on all the cosmological parameters from the simulation particles, using the $k\nn$ analysis and the $\xi(r)$ analysis is presented in Fig. \ref{fig:part_constraints}. Here we use the full data vector outlined in Sec. \ref{sec:fisher}. Since the fiducial cosmology has $M_\nu=0$, and negative $M_\nu$ values are unphysical, we follow the example in \cite{2019arXiv191111158U}, and plot constraints on $\Delta M_\nu$ instead, where $\Delta M_\nu$ is the change in the total neutrino mass from the fiducial value. The $1$-$\sigma$ constraints on individual parameters is listed in Table \ref{tab:part_constraints}. For all the cosmological parameters, we find that the $k\nn$ analysis improves the constraints by almost a factor of $2$ over the $\xi(r)$ analysis. While some of the degeneracy directions between pairs of parameters are somewhat different between the two analyses, we find that both  are affected by a strong $M_\nu-\sigma_8$ degeneracy, as can be expected when utilizing information only from the relatively small scales used in these analyses. Note that our choice of leaving out the larger scales, including the BAO peak from the analysis also affects the shape and size of the other contours and the degeneracy directions compared to other works. We conclude that on the scales between $10\hmpc$ and $40\hmpc$, using only a two-point function analysis fails to capture at least half the total information about cosmological parameters that is available on these scales. The $k\nn$ analysis, on the other hand, is sensitive to higher order clustering, and is much more sensitive to changes in the cosmological parameters. In Appendix \ref{sec:downsampling}, we show that the results presented here are consistent with those obtained from a slightly different formalism where only the nearest neighbor distribution is used, but with different mean densities for the data points.

\begin{table}
	\centering
	\caption  {$1$-$\sigma$ constraints on cosmological parameters, from the $k\nn$ and $\xi(r)$ analysis of simulation particles using scales in the range $10\hmpc$ to $40\hmpc$.}
	\label{tab:part_constraints}
	\begin{tabular}{c|c|c}
		\hline
		\hline
		${\rm Parameter}$ & $\sigma_{k\nn}$ & $\sigma_{\xi(r)}$\\
		\hline
		\hline
		$\Omega_b$ & 0.0023 & 0.0069 \\
		$n_s$ & 0.0197 & 0.0521 \\
		$h$ & 0.0224 & 0.0450 \\
		$\Omega_m$ & 0.0105 & 0.0201 \\
		$\sigma_8$ & 0.0055 & 0.0156 \\
		$\Delta M_\nu$ & 0.0792 & 0.173 \\
		$w$ & 0.0612 & 0.115 \\
		\hline		
	\end{tabular}
\end{table}

\subsection{Simulation Halos}
\label{sec:cosmo_constraints_halos}

We now turn to the parameter constraints from the analysis of halos in the simulation.Throughout this section, we will consider fixed number density samples across different cosmologies. Specifically, we focus on the clustering statistics of the $10^5$ most massive halos in each simulation volume. The halos are identified using an FoF algorithm from the simulation particles. Even for comsologies with neutrino particles in the simulation, the halo finding is run only on the CDM particles. At $z=0$, this number cut corresponds to mass cuts around $5\times10^{13}M_\odot/h$, and at $z=0.5$, corresponds to mass cuts around $3\times 10^{13}M_\odot/h$. As discussed previously, the $k\nn$ statistics is sensitive to the mean number density, and using a fixed number density ensures that differences in the $k\nn$ statistics at different cosmologies arise only from changes in the underlying clustering. In the language of the Fisher formalism, this analysis quantifies the response of the clustering observables --- $k\nn$ distributions or $\xi(r)$ measurements --- constructed from the \textit{$10^5$ most massive halos} in a $(1\hgpc)^3$ volume, to a change in the cosmological parameters, and then converts the amplitude of the response into parameter constraints. Once again, we only use scales in the range $10\hmpc$ to $40\hmpc$ in our analysis.

\begin{figure*}
	\includegraphics[width=0.9\textwidth]{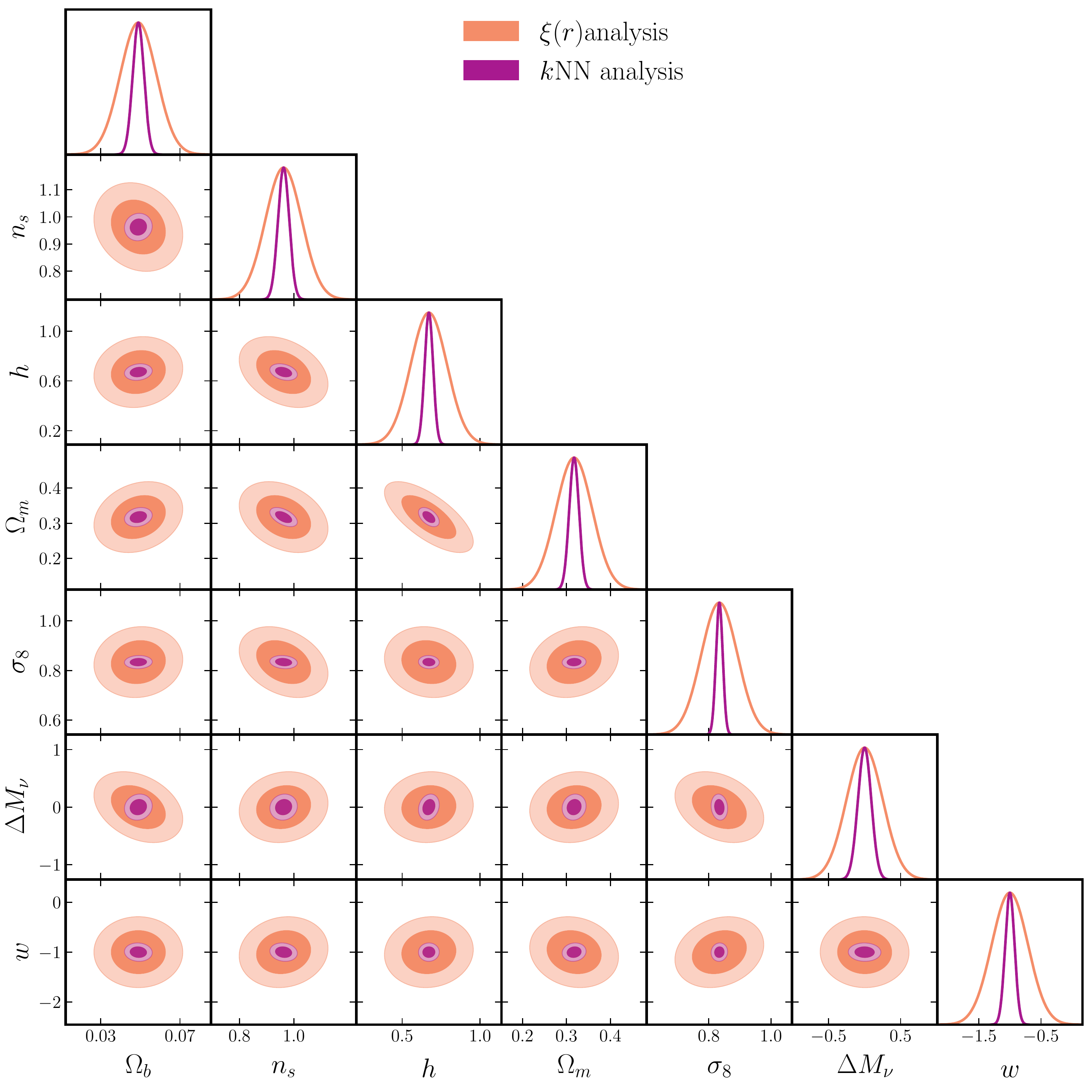}
	\caption{Constraints on the cosmological parameters derived from the Fisher analysis of the clustering of the $10^5$ most massive halos at different cosmologies, combining information from $z=0$ and $z=0.5$, and using scales in the range $10\hmpc$ to $40\hmpc$. The constraints from the $k\nn$ analysis are much tighter than those from the $\xi(r)$ analysis. The values of the constraints on individual parameters are listed in Table \ref{tab:halo_constraints}.}
	\label{fig:halo_constraints}
\end{figure*}

\begin{table}
	\centering
	\caption{$1$-$\sigma$ constraints on cosmological parameters, from the $k\nn$ and $\xi(r)$ analysis of the real space clustering of the $10^5$ most massive halos in the simulation volume of $(1\hgpc)^3$, using scales in the range $10\hmpc$ to $40\hmpc$.}
	\label{tab:halo_constraints}
	\begin{tabular}{c|c|c}
		\hline
		\hline
		${\rm Parameter}$ & $\sigma_{k\nn}$ & $\sigma_{\xi(r)}$\\
		\hline
		\hline
		$\Omega_b$ & 0.0029 & 0.0092 \\
		$n_s$ & 0.0206 & 0.0667 \\
		$h$ & 0.0273 & 0.1165 \\
		$\Omega_m$ & 0.0111 & 0.0413 \\
		$\sigma_8$ & 0.0108 & 0.0584 \\
		$\Delta M_\nu$ & 0.0925 & 0.2520 \\
		$w$ & 0.0756 & 0.2916 \\
		\hline		
	\end{tabular}
\end{table}

\begin{figure*}
	\includegraphics[width=0.9\textwidth]{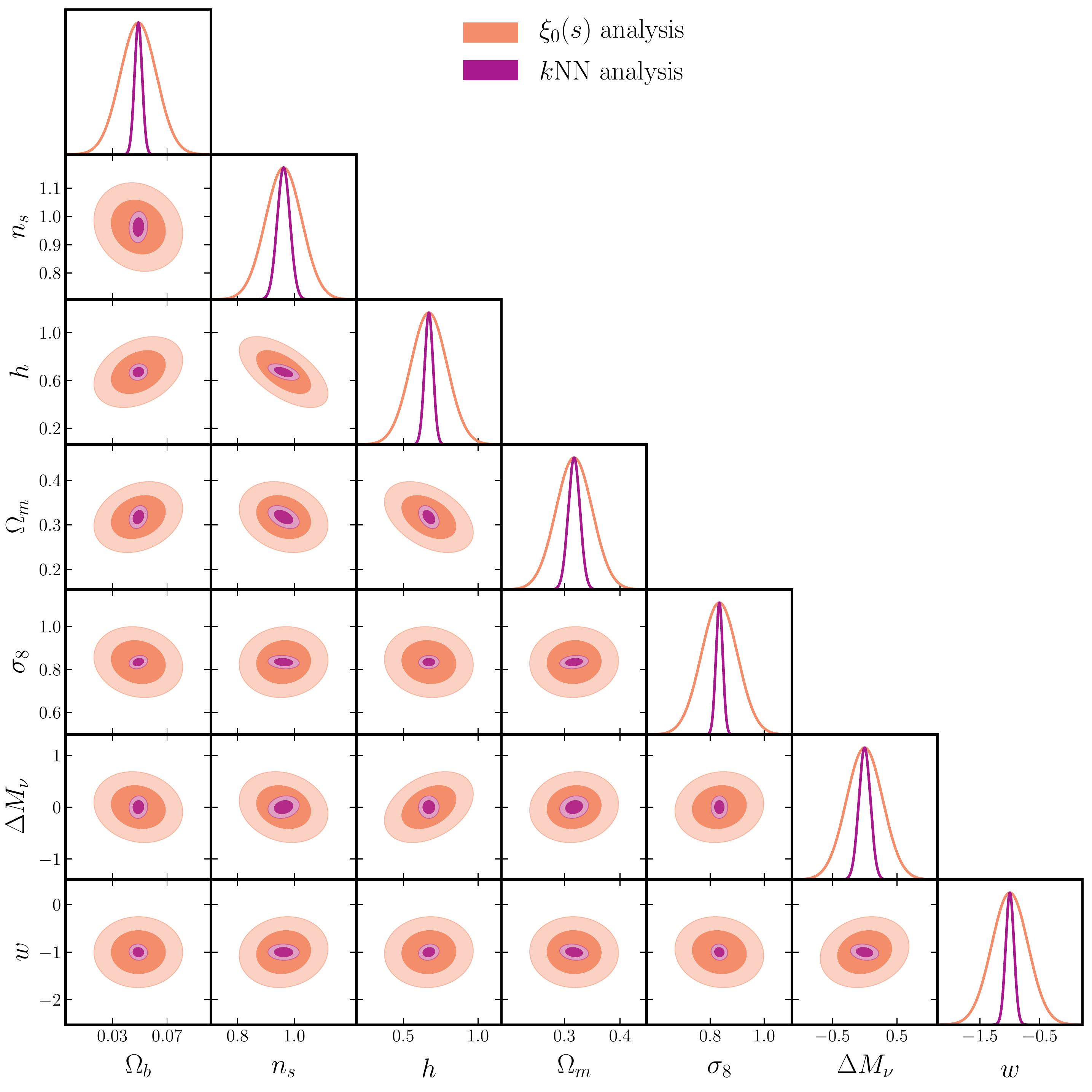}
	\caption{Constraints on the cosmological parameters derived from the Fisher analysis of the monopole of clustering of the $10^5$ most massive halos in redshift space, combining information from $z=0$ and $z=0.5$, and using scales in the range $10\hmpc$ to $40\hmpc$. Similar to Fig. \ref{fig:halo_constraints}, the constraints from the $k\nn$ analysis are much tighter than those from the $\xi(r)$ analysis. }
	\label{fig:RSD_constraints}
\end{figure*}

If the two-point function is used as the summary statistic, parameter constraints degrade significantly when analyzing halo clustering compared to matter clustering. This is due to halo bias, which, in general, is unknown. Since the effect of the linear bias term on the two-point function, especially on large scales, is degenerate with a number of cosmological parameters, like $\sigma_8$ and $M_\nu$, marginalizing over the unknown bias term relaxes the constraints on these. It is only possible to break the degeneracy by either using other observables (usually lensing) which have a different dependence on bias, or by using information from smaller scales, where the linear bias approximation breaks down. Note that, in general, it is not necessary for the linear bias approximation to break down at the same scale as where the underlying clustering becomes non-Gaussian \citep{2020MNRAS.492.5754M}.

We have already demonstrated in Sec. \ref{sec:b_s8} that the $k\nn$ statistics can break the degeneracy between linear bias and the amplitude of clustering of the underlying field. The $k\nn$ statistics were able to distinguish between two halo samples where the underlying clustering of the matter field were different, but the bias (or mass cut) adjusted such that they have the same two-point clustering signal. It is therefore, possible, that more information about the cosmological parameters is retained in the $k\nn$ analysis of massive halos, compared to a $\xi(r)$ analysis.

This is indeed what we find from the Fisher analysis on the clustering of halos in real space, and the results are presented in Fig. \ref{fig:halo_constraints}. The values of the $1$-$\sigma$ constraints on individual parameters are tabulated in Table \ref{tab:halo_constraints}. Unsurprisingly, we find that for $\xi(r)$, the constraints are significantly relaxed when compared to the matter clustering case in Fig. \ref{fig:part_constraints} and Table \ref{tab:part_constraints}. This is most significantly apparent for $\sigma_8$, where the constraints degrade by almost a factor of $8$.  On the other hand, the constraints yielded by the $k\nn$ analysis are much closer to those from the analysis of the matter field. Even though the constraint on $\sigma_8$ is less stringent than for the matter field analysis, it is still tighter than the $\xi(r)$ constraint by a factor of $\sim 5$.

Therefore, the constraints on all cosmological parameters are significantly improved when using the $k\nn$ statistics, and the improvement over the two-point analysis using the same scales is larger when considering the $10^5$ most massive halos than when considering the clustering of the underlying matter field. This has many potential implications for the analysis of galaxy clustering in cosmological surveys, both photometric, and spectroscopic.

While the analysis presented above considers the clustering of halos in real space, spectroscopic surveys generally measure the clustering of galaxies in redshift space. It is, therefore, worthwhile to understand, if the projection to redshift space affects the relative improvements on parameter constraints from the $k\nn$ analysis over the $\xi(r)$ analysis. To convert the real space positions $\mbr $ of the halos in the simulation volume to redshift space positions $\mathbf s$, we simply use
\eq{redshift_space}{\mathbf s=\mbr + \frac{1+z}{H(z)}v_x\hat x\,,}
where $H(z)$ is the Hubble parameter at redshift $z$. In the above equation, we have assumed that the $\hat x$ direction is the line-of-sight direction. 

Because the line of sight projection breaks the rotational symmetry, clustering in redshift is no longer isotropic, and it is common to include the quadrupole and the hexadecapole of the correlation function in addition to the monopole. To keep the analysis simple, we focus here only on the monopole for the two point correlation function, and compare it to the $k\nn$ statistics in redshift space. To construct the CDFs for the $k\nn$ distributions, we proceed exactly as for the real space calculations, once all the halo positions have been transformed according to Eq. \ref{eq:redshift_space}. 

The results of the analysis are plotted in Fig. \ref{fig:RSD_constraints}. Once again, we find that using the $k\nn$ statistics yields much tighter constraints on the cosmological parameters compared to the two-point function analysis. The gain is very similar to those that were obtained in the real space analysis above. Adding in the quadrupole and hexadecapole two-point functions should add more constraining power to the two-point analysis, but it should be noted that the $k\nn$ statistics can also be expanded in the presence of a special line-of-sight direction. For example, the nearest neighbor distributions along the LOS direction and the perpendicular directions can be computed separately - the differences in the distribution along individual directions arise exactly from the redshift projection. We will explore these issues in detail in future work.




\section{Summary and Future Directions}
\label{sec:conclusions}

In this paper we have presented the nearest neighbor distributions ($k\nn$-$\cdf$), the empirical cumulative distribution function of distances from a set of volume-filling Poisson distributed random points to the $k$--nearest data points, as a set of new summary statistics that can be used in cosmological clustering analysis. We have shown how these are related to the counts-in-cell distributions, and to the various $N$-point correlation functions of the clustering of a set of data points. We have outlined how these distributions can be efficiently computed over a range of scales using the data points, or a subsample of them, and a set of volume filling randoms. We have applied the $k\nn$-$\cdf$ formalism to the clustering of tracers drawn from two different continuous distributions, \textit{i.e.} the Poisson sampling of an uniform field, and a Gaussian field. We also demonstrate its application to data from a cosmological simulation - both for simulation particles, and dark matter halos identified in the simulations. We have explored how the nearest neighbor distributions can break the degeneracy between linear halo bias and the amplitude of clustering of the underlying matter distribution, which is a major limitation of the two-point statistics on large scales. Then, employing the Fisher matrix formalism, we have quantified the constraints that can be placed on various cosmological parameters by using the $k\nn$-$\cdf$ as the summary statistics. Using simulation particles, and scales between $10\hmpc$ and $40\hmpc$, we demonstrate that the use of $k\nn$-$\cdf$ distributions improve the constraints by a factor of $\sim 2$ over those from the two-point correlation function analysis. We have demonstrated that these gains are even larger, roughly a factor of $4$, when applied to the analysis of the clustering of halos in real space with a fixed number density, and that these gains are expected to persist even after projecting to redshift space, making it a promising tool for analyzing cosmological data.

In addition to the sensitivity of these $k\nn$-$\cdf$ statistics to changes in the underlying cosmology, the fact that they can computed extremely efficiently, is highly advantageous. Other methods which aim to harness the extra constraining power, beyond the two-point information, usually are computationally expensive. For example, measuring the bispectrum is much more expensive than the power spectrum, while the trispectrum is even more so. Other approaches, such as measuring the density PDF at different scales usually require multiple Fourier transforms, at least two for each scale at which the PDF is to be computed.
For the $k\nn$-$\cdf$s on the other hand, search times for the $k$-th nearest neighbor does not scale with $k$ once the tree has been constructed, and so, information about higher order cumulants can be accessed at negligible additional computational cost. Further, a single computation is sufficient to provide information on a wide range of scales, and a range of values for $k$. While we have focused in this paper on clustering in $3$ dimensions, the formalism presented can easily be applied to non-Gaussian data in $2$ dimensions, where the calculations are computationally even cheaper due to the reduction in the dimensionality. This framework is therefore also well suited to the study the tomographic clustering of galaxies in a photometric survey.

For distributions with only a finite number of connected $N$-point functions, say $m$, measuring only the $m$ nearest neighbor distributions is sufficient to capture the full clustering. For cosmological applications, one does not \textit{a priori} know the value of $m$, or if $m$ is even finite, but as has been demonstrated, we exhaust most of the independent information by the time, \textit{e.g}, the $8$-th nearest neighbor distribution is used in the analysis. This value is of course dependent on the choice of scale cuts used in the analysis, and if smaller scales are used, it is likely that a few more nearest neighbor measurements will be needed to capture all the information. As we have pointed out, we have been conservative in our choice of scale cuts based on the resolution of the simulations used in the analysis, and including smaller scales will yield even greater improvements in parameter constraints over the two-point analysis.

While the $\nn$-$\cdf$ formalism lends itself most naturally to the analysis of clustering of discrete points, we have demonstrated that it can also be used to study statistical properties of continuous fields. The matter density field has been analyzed by downsampling the simulation particles. Similarly, statistical properties of other continuous fields, such as the convergence fields for weak lensing measurements, can also be analyzed in the same way by appropriately sampling the field. The sampling rate is determined by the scales of interest, such that different distribution functions are robustly measured, and have to be tailored for the application at hand. However, once the sampling rate has been fixed, all other aspects of the analysis presented here can be carried through without major changes in the workflow. 

In the current work, we have used the Fisher formalism to demonstrate the improvement in constraining power when $\nn$-$\cdf$s are used, the Fisher formalism is only reliable when certain approximations are valid. Further, note that the empirical CDF is composed of $\sim 10^6$ measurements, but in the Fisher analysis, we only use the $\cdf$s interpolated to $16$ radial scales. This is needed such that the Hartlap factor in Eq. \ref{eq:Hartlap}, given the number of simulations, remains close to unity, and that the Fisher constraints are reliable. Any analysis of actual data will likely demand more sophisticated statistical techniques. Since the actual measurements in the nearest neighbor calculations are those of Cumulative Distribution Functions, the Kolmogorov--Smirnov (KS) test \citep{smirnov1948} lends itself quite naturally as an alternative. Another advantage of the KS test, when a large number of measurements is at hand is that the test has more statistical power when the empirical CDF is sampled at higher rates. Other measures such as the Kullback--Leibler divergence \citep{kullback1951} can also potentially be applied in conjunction with the $\nn$-$\cdf$ measurements, and these issues will be explored in more detail in future work.

We have focused here on the improvement in parameter constraints for samples where the number densities are high enough for the two-point function to be computed using standard methods without being strongly affected by shot noise considerations. However, as demonstrated in Sec.\ref{sec:Gaussian}, the $\nn$-$\cdf$ method also allows for the measurement of the clustering signal even on scales where shot noise is expected to dominate, i.e. on scales smaller than the mean inter-particle separations. The stronger the clustering, the easier it is to robustly detect the clustering at a fixed mean number density. In cosmology, interesting samples with low number density, such as the most massive clusters, also tend to be highly clustered with respect to the underlying matter field, implying that the $\nn$-$\cdf$ method can be applied to the study of the clustering of extremely rare objects such as the most massive galaxy clusters, which are usually completely noise-dominated when computed using the standard methods. 

Another possible application of nearest neighbor statistics is as a test of non-Gaussianity in the clustering of objects. If the number density of a sample is well known, and the variance of the distribution can be computed as a function of scale, the nearest neighbor measurements can be used to check if the clustering matches that of a fully Gaussian distribution, as discussed in Sec. \ref{sec:Gaussian}. Any departures from the expressions for the nearest neighbor distributions derived in Sec. \ref{sec:Gaussian} can be interpreted as the presence of non-Gaussian terms. On small scales, this can help test the range of validity of analysis methods that rely on assumptions of Gaussianity, while on larger scales such measures could be relevant for the search for primordial non-Gaussianity in Large Scale Structure.

At a fixed cosmology, the $\nn$-$\cdf$ measurements can be used to study the clustering of different halo samples. One such application is in the context of halo and galaxy assembly bias. For example, different galaxy samples targeted in spectroscopic surveys are known to occupy different parts of the cosmic web \citep{2018MNRAS.475.2530O,2019MNRAS.483.4501A}. Under these conditions, a single linear bias parameter in the measured two-point correlation function does not reflect the full difference in the clustering \citep{2019MNRAS.488.3541W}. The nearest neighbor framework offers a way to capture these differences more clearly. On a related note, adding $\nn$-$\cdf$ measurements to the data vectors used in studies of the galaxy-halo connection can add to the power of these models, and can improve constraints on cosmological parameters even after marginalizing over all the galaxy-halo connection parameters.

\section*{Acknowledgements}

This work was supported by the U.S. Department of Energy SLAC
Contract No. DE-AC02-76SF00515. The authors thank the referee, Mark Neyrinck, for comments which helped improve the paper. The authors would like to thank Susmita Adhikari and Michael Kopp for useful discussions, and Daniel Gruen, Michael Kopp, Johannes Lange, Jeff Scargle, Istvan Szapudi, Cora Uhlemann, and Francisco Villaescusa-Navarro for comments on an earlier version of the manuscript. Some of the computing for this project was performed on the Sherlock cluster. The authors would like to thank Stanford University and the Stanford Research Computing Center for providing computational resources and support that contributed to these research results. The \textsc{Pylians3}\footnote{https://github.com/franciscovillaescusa/Pylians3} analysis library  was used extensively in this paper. We acknowledge the use of the \textsc{GetDist}\footnote{https://getdist.readthedocs.io/en/latest/} \citep{2019arXiv191013970L} software for plotting.

\section*{Data Availability}

The simulation data used in this paper is publicly available at \url{https://github.com/franciscovillaescusa/Quijote-simulations}. Additional data is available on reasonable request.



\bibliographystyle{mnras}
\bibliography{ref} 




\appendix


\section{Derivation of the generating function for discrete counts}
\label{sec:derivation}
We derive the expression for Eq. \ref{eq:generating_function} by summarizing the arguments presented in \cite{Szapudi1993}.
The generating functional for a continuous field $\rho(\mbr)$ can be written as an integral over all field configurations of $\rho$:
\begin{align}
	\mathcal Z[J] = \int \big[D\rho(\mbr)\big]P\big[\rho(\mbr)\big] \exp\Bigg[i  \int d^3 \mbr \rho(\mbr) J (\mbr)\Bigg] \, 
	\label{eq:generating_functional}
\end{align}
where $\big[D\rho(\mbr)\big]$ represents the functional integral over field configurations, and $P\big[\rho(\mbr)\big]$ represents the probability of a specific field configuration. In this context, the field $\rho$ will generally represent a density field. For a mean density $\bar \rho$, the cumulants of the distribution are related to the generating functional through the following equation:
\begin{align}
	\bar \rho^k\xi^{(k)}(\mathbf r_1...\mathbf r_k) = \frac{(-i)^k\delta^k \big(\ln \mathcal Z[J]\big)}{\delta J(\mathbf r_1)...\delta J(\mathbf r_k)} \,.
	\label{eq:cumulant_definition}
\end{align}
Eq. \ref{eq:cumulant_definition} can be inverted to express the generating functional itself in terms of the cumulants:
\begin{align}
	\mathcal Z[J] = \exp\Bigg[\sum_{k=0}^\infty \frac{(i\bar \rho)^k}{k!}& \int d^3 \mathbf r_1...d^3 \mathbf r_k \xi^{(k)}\left(\mbr_1...\mbr_k\right) \times \nonumber \\
	 &  \qquad \qquad J(\mbr_1)...J(\mbr_k)\Bigg] \, .
	 \label{eq:functional_cumulants}
\end{align}
We now consider a set of points generated by a local Poisson process where the number of points generated in a volume $V$ around $\mbr$ depends only on the local integrated density over that volume, $\rho_V(\mbr)$:
\eq{smoothed_density}{\mathcal M_V(\mathbf r) = \int_V d^3 \mathbf r' \rho(\mathbf r')W(\mathbf r, \mathbf r')\,,}
where $W(\mbr, \mbr ')$ defines the window function for the smoothing procedure.
Therefore, at a given point $\mbr$ with integrated density $\mathcal M_V(\mbr)$, the probability of finding $k$ points within radius $r$ is 
\begin{align}
	P_{k|\mathcal M_V} = \frac{(\mathcal M_V/m)^k}{k!}\exp\big[-\mathcal M_V/m\big] \, ,
\end{align}
where $m$ is the ``mass" associated with each particle. While this has a relatively straightforward meaning when applied to simulation particles, for halos, this can be thought of as a  normalization factor. The overall probability of finding $k$ points within volume $V$ of $\mbr$ needs to average over all possible field configurations:
\eq{overal_prob}{P_{k|V} = \Bigg\langle \frac{(\mathcal M_V/m)^k}{k!}\exp\big[-\mathcal M_V/m\big] \Bigg\rangle \, .}

The generating function for the discrete distribution is then defined as 
\eq{discrete_generating_function}{P(z|V) &= \sum_{k=0}^\infty P_{k|V} z^k  = \sum_{k=0}^\infty \Bigg\langle \frac{(\mathcal M_V/m)^k}{k!}\exp \big[-\mathcal M_V/m \big] \Bigg\rangle z^k \nonumber \\
& = \Big\langle \exp\big[\mathcal M_V(z-1)/m\big]\Big\rangle \, .}
Note that in the last line of Eq. \ref{eq:discrete_generating_function}, all quantities are in terms of the continuous variables. Therefore this expectation value can be computed through the functional integral over all configurations of the continuous field $\rho$:
\eq{generating_cont_to_discrete}{P(z|V) =& \int \big[D\rho(\mbr)\big]P\big[\rho(\mbr)\big] \times \nonumber \\ & \qquad \exp\Bigg[ \frac{(z-1)}{m} \int d^3 \mbr  \rho(\mbr ) W (\mbr, \mbr ') \Bigg] \, .}
Therefore, Eqs. \ref{eq:generating_functional} and \ref{eq:generating_cont_to_discrete} match each other when $J(\mbr) = W(\mbr, \mbr ')(z-1)/(i m)$. We now use the fact that the RHS of Eqs. \ref{eq:generating_functional} and \ref{eq:functional_cumulants} are equivalent, to write
\eq{}{P(z|V) = \exp\Bigg[\sum_{k=0}^\infty \frac{\big(\bar n(z-1)\big)^k}{k!}& \int d^3 \mathbf r_1...d^3 \mathbf r_k \xi^{(k)}\left(\mbr_1',...,\mbr_k '\right) \times \nonumber \\
&  \qquad  W(\mbr_1,\mbr_1 ')...W(\mbr_k, \mbr_k ')\Bigg] \, ,}
where $\bar n = \bar \rho /m$. 
For the special case $W(\mbr, \mbr ') = 1$ only when both points are in the volume considered, and $0$ everywhere else, such as in top-hat smoothing, we get the expression in Eq. \ref{eq:generating_function},
\eq{}{P(z|V) &= \exp\Bigg[\sum_{k=1}^\infty \frac{\bar n ^k (z-1)^k }{k!} \times  \nonumber \\ 
& \quad \qquad \int _V ... \int_V d^3 \mathbf r_1 ... d^3 \mathbf r_k \xi^{(k)} \left(\mathbf r_1,...,\mathbf r_k\right)\Bigg] \, .}

\section{Cumulants}
\label{sec:cumulants}

The moment generating function for the discrete distribution is given by 
\begin{align}
	M(t|V) &= \sum_{k=0}^\infty \exp\big[t k\big]P_{k|V} \nonumber\\
	&= \exp\Bigg[\sum_{k=1}^\infty \frac{\bar n ^k (e^t-1)^k }{k!} \times  \nonumber \\ 
	& \quad \qquad \int _V ... \int_V d^3 \mathbf r_1 ... d^3 \mathbf r_k \xi^{(k)} (\mathbf r_1,...,\mathbf r_k)\Bigg] \, .
\end{align}
The $k$th cumulant $\mathcal C_k$ is given by
\eq{}{\mathcal C_k = \Bigg[\bigg(\frac{d}{dt}\bigg)^k\ln \bigg(M(t|V) \bigg)\Bigg]_{t=0} \, .} 
The first and second cumulants therefore work out to be 
\eq{}{\mathcal C_1 &= \bar n V \\
	\mathcal C_2 &= \bar n V + \bar n^2 \int_V\int_V d^3 \mbr_1 d^3 \mbr_2 \xi^{(2)}(\mbr_1, \mbr_2)\, ,}
where $\xi^{(2)}$ is the usual two-point correlation function. Note that $\xi^{(2)}(\mbr_1, \mbr_2) = \xi^{(2)}\left(r = |\mbr_1 - \mbr_2|\right)$ due to isotropy. For the special case when the volume $V$ is associated with a sphere of radius $r_0$, we can write $\mathcal C_2$ as a one dimensional integral over $\xi^{(2)}(r)$ with some weight function $W(r|r_0)$:
\eq{C2_from_xi}{\mathcal C_2 = \bar n V + \bar n^2 \int_0^{2r_0} dr \, r^2 \xi^{(2)}(r) W(r|r_0)\, ,}
where 
\eq{}{W(r|r_0) = 8 \pi^2 \left(\frac{1}{12}(2r_0-r)^3 + \frac {1}{8} r(2r_0-r)^2\right)\,.}
The second moment or variance can also be expressed in terms of the power spectrum, $P(k)$, smoothed on scale $r$ corresponding to volume $V$:
\eq{variance}{\mathcal C_2 =  \bar n V +\frac{(\bar n V)^2 }{2\pi^2}\int d k \, k^2 P(k) W^2(kr) = \bar n V + \bar n^2 V^2 \sigma_V^2 \, .}
$W(kr)$ represents the Fourier transform of the tophat window function for a filter with radius $r$:
\eq{}{W(kr) = \frac{3}{(kr)^3}\bigg(\sin \big(kr\big) - \big(kr\big)\cos\big(kr\big)\bigg)\,.}
Note that Eq. \ref{eq:variance} is true for any distribution - the distribution need not be a Gaussian. Therefore, $\xi(r)$ or $P(k)$, along with the knowledge of the mean number density, encodes information about the second cumulant of the distribution, even in the nonlinear regime.

\begin{figure}
	\includegraphics[width=\columnwidth]{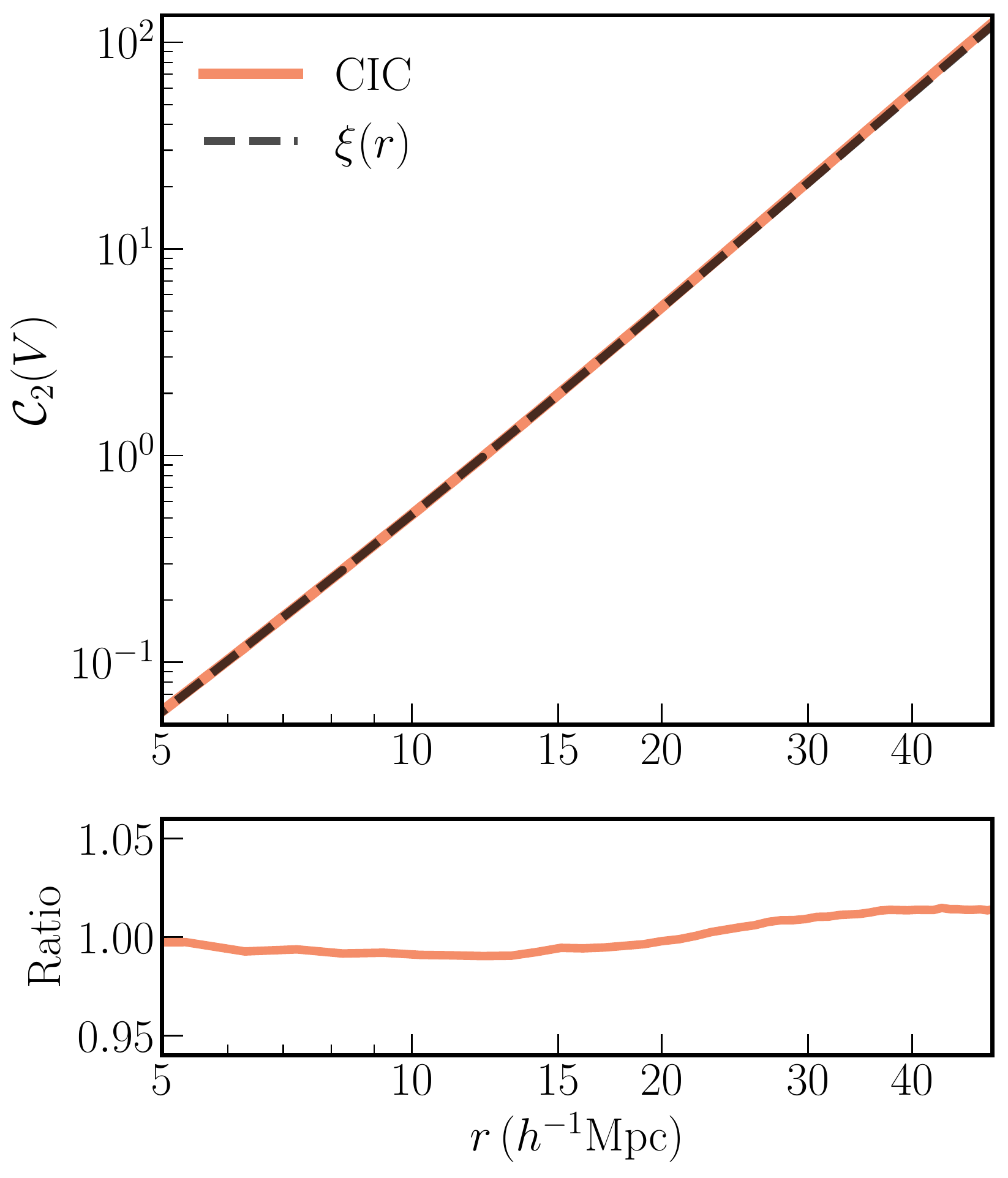}
    \caption{\textit{Top}: Second cumulant $\mathcal C_2$, as a function of scale, for the distribution of $10^5$ simulation particles in a $(1\hgpc)^3$ volume at $z=0$. The solid line represents the result for $C_2$ computed using the CIC distributions, according to Eq. \ref{eq:cumulant_cic}. The CIC distributions themselves have been computed from the $k\nn$ distributions, according to Eq.\ref{eq:cic_knn}. The dashed line represents $\mathcal C_2$ computed from the measured $\xi(r)$ of the full set of simulation particles, using Eq. \ref{eq:C2_from_xi}. \textit{Bottom}: Ratio of the two measurements from the top panel.}
    \label{fig:C2}
\end{figure}

\begin{figure}
	\includegraphics[width=\columnwidth]{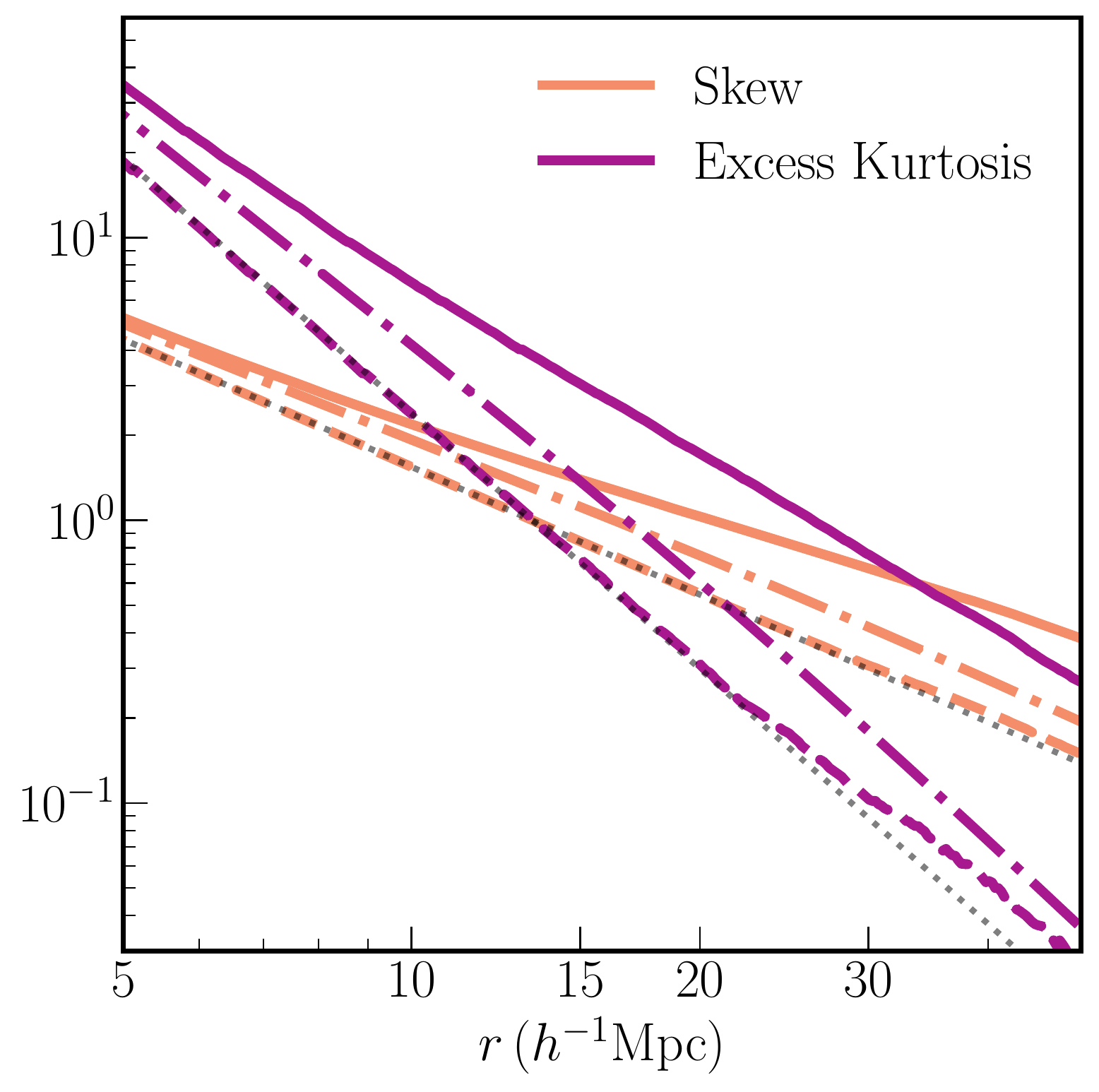}
    \caption{Solid lines represent the skew and excess kurtosis (as a function of scale) of the distribution of $10^5$ simulation particles in a $(1\hgpc)^3$ volume at $z=0$. The dot-dashed lines represent the expected skew and kurtosis values for a set of discrete points with the same number density, and the same two point function, but no higher connected $n-$ point functions. The dashed lines represent the measurements for a set of Poisson distributed points with the same number density. The thin dotted lines represent the analytic predictions for the skew and excess kurtosis of the Poisson distribution.}
    \label{fig:higher_cumulants}
\end{figure}

The second cumulant can also be computed from the CIC distributions:
\eq{cumulant_cic}{\mathcal C_2 (V) = \frac{\sum_{k=0}^\infty k^2 P_{k|V}}{\sum_{k=0}^\infty P_{k|V}}-\Bigg(\frac{\sum_{k=0}^\infty k P_{k|V}}{\sum_{k=0}^\infty P_{k|V}}\Bigg)^2\,.}
While the sum formally runs from $0$ to $\infty$, in practice, for any realistic distribution function, a finite number of terms is sufficient to determine $\mathcal C_2$ accurately. Fig. \ref{fig:C2} shows the measurement of $\mathcal C_2(V)$ for a set of $10^5$ simulation particles in a $(1\hgpc)^3$ volume at $z=0$. The solid line represents the measurement from the data using Eq. \ref{eq:cumulant_cic}. The CIC measurements themselves are derived from the $k\nn$ measurements through Eq. \ref{eq:cic_knn}. The dotted line represents the measurement of $\mathcal C_2$ using Eq. \ref{eq:C2_from_xi}, where $\xi(r)$ was computed from all $512^3$ particles in the simulation, but $\bar n$ was set to match the number density considered for the CIC measurement. The two measurements produce consistent results at the $\sim 1\%$ level, even though we use only $200$ nearest neighbor distributions to compute the solid curve - where, formally, the sums in Eq. \ref{eq:cumulant_cic} run from $0$ to $\infty$. 

Similarly, higher cumulants can also be obtained directly from the $\nn$-$\cdf$ by taking higher moments of the $P_{k|V}$, as in Eq. \ref{eq:cumulant_cic}. These cumulants are directly related to the connected higher $N$-point functions and their Fourier equivalents, just as the second cumulant is related to the two-point function or $P(k)$. We plot the skewness and excess kurtosis, which are related to the third and fourth cumulants, of the same sample of particles from the simulations using the solid lines in Fig. \ref{fig:higher_cumulants}. The dot-dashed lines are the expected skew and kurtosis for a set of points with the same number density, and the same two-point function, but with no higher connected $n-point$ functions. The dashed lines of the same color represent the measurements of the the same quantities from a set of Poisson distributed points with the same mean density. The dotted lines represent the analytic predictions for the skew and excess kurtosis for the Poisson distribution. We find very good agreement between the measurements and the predicted values for the Poisson distribution, while the measurements from the cosmological distribution (solid lines) are clearly different from that of the Poisson distribution on all scales displayed on the plot. This is true even in scales below the mean inter-particle separation, where shot noise could potentially be important. Further, comparing the solid lines and the dot-dashed lines on the plot illustrates the fact that this measurement is indeed sensitive to the presence of higher cumulants in the data. Once again, we have terminated the calculation at $200$ nearest neighbors, just as we did for the calculation of the second cumulant. It is also worth noting that the time taken to compute quantities related to the second, third, fourth, and potentially higher cumulants is the same - \textit{i.e.} once the $\nn$-$\cdf$ distributions for the 200 neighbors has been computed using the tree, the rest of the calculation is computationally trivial, irrespective of the order of the cumulant. Of course, sample variance limitations of the nature discussed in the text imply that the lower cumulants are more robustly measured over a larger range of scales. This can be seen in the departure of the measurement and analytic expectation of the excess kurtosis, or the fourth cumulant, of the Poisson distribution on large scales in Fig. \ref{fig:higher_cumulants}.

\section{Data downsampling}
\label{sec:downsampling}

\begin{figure*}
	\includegraphics[width=0.9\textwidth]{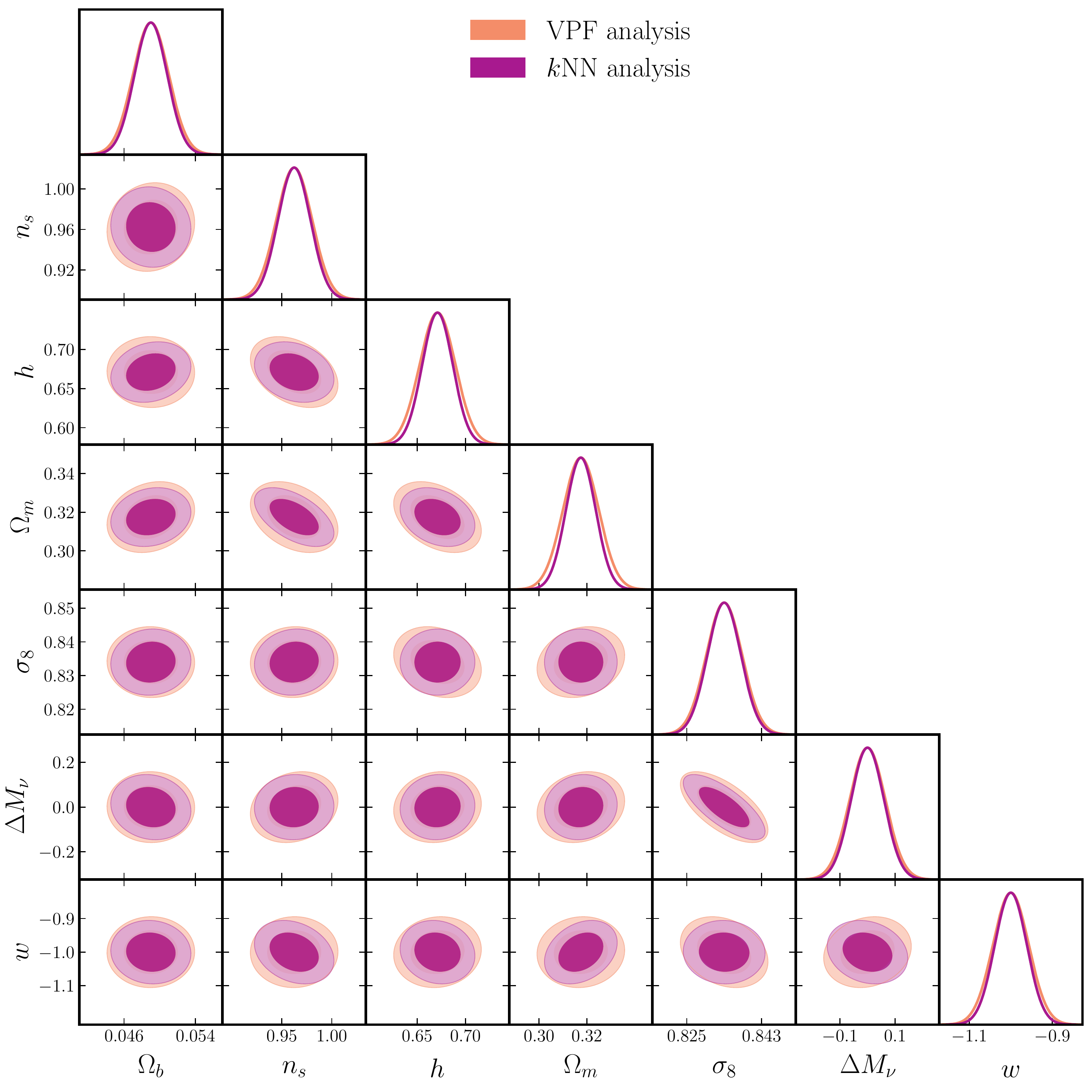}
	\caption{Fisher constraints on cosmological parameters from the $k\nn$ analysis in Fig. \ref{fig:part_constraints} and the VPF analysis outlined above. The two formalisms yield very similar constraints on all the cosmological parameters.}
	\label{fig:kNN_vs_VPF}
\end{figure*}

In the paper introducing the Void Probability Function, \cite{White1979} showed that the VPF itself can serve as the generating function for the $P_{k|V}$ distribution. In this formalism, however, the derivatives had to be taken with respect to the mean number density $\bar n$, unlike Eq. \ref{eq:counts}, where the derivatives are taken with respect to the dummy variable $z$. As we have shown earlier,
\eq{vpf_1nn}{{\rm VPF}(r) = 1-\cdf_{1\nn}(r)\,.}
In theory, therefore, the information in the $k$-th neighbor distribution can also be accessed by computing the nearest neighbor distribution at a different mean number density.

We now compare the actual parameter constraints from the Fisher analysis of $k\nn$ distributions for simulation particles, presented in Sec. \ref{sec:cosmo_constraints_particles}, with those that can be obtained by computing the VPF at different number densities. In order to do this, we compute the VPF for subsamples of the simulation particles with $\bar n = \{1\times10^{-4},0.5\times 10^{-4}, 0.25\times 10^{-4}, 0.125\times 10^{-4}\}(\hmpc)^{-3}$, using Eq. \ref{eq:vpf_1nn}. Note that the $1\times 10^{-4}(\hmpc)^{-3}$ is the mean number density used in Sec. \ref{sec:cosmo_constraints_particles}. The other values of $\bar n$ are also chosen in a way that they correspond to the $k\nn$ measurements presented there. The data vector is created by appending the VPF measurements at $16$ logarithmically spaced scales for each $\bar n$ in the scale range of $10\hmpc$ to $40\hmpc$. We once again use the relevant Erlang distribution to ensure that the scales do not include measurements of the VPF deep into the tails. As earlier, we combine measurements from $z=0$ and $z=0.5$.

The results of the Fisher analysis are presented in Fig. \ref{fig:kNN_vs_VPF}, where the constraints from the VPF formalism above is contrasted with the constraints from the $k\nn$ analysis presented in Fig. \ref{fig:part_constraints} and Table \ref{tab:part_constraints}. The final constraints on individual parameters are very similar, though some of the degeneracy directions are slightly different. The agreement in the constraints hold true even though the different derivatives and the covariance matrices between the two analyses look very different. This is a useful check on the robustness of the constraints obtained from the $k\nn$ distributions, as a well as a test of the understanding of the underlying statistical methods.

\bsp	
\label{lastpage}
\end{document}